\PassOptionsToPackage{unicode}{hyperref}
\PassOptionsToPackage{hyphens}{url}
\PassOptionsToPackage{dvipsnames,svgnames,x11names}{xcolor}
\documentclass[
  12pt]{article}

\usepackage{amsmath,amssymb}
\usepackage{iftex}
\ifPDFTeX
  \usepackage[T1]{fontenc}
  \usepackage[utf8]{inputenc}
  \usepackage{textcomp} % provide euro and other symbols
\else % if luatex or xetex
  \usepackage{unicode-math}
  \defaultfontfeatures{Scale=MatchLowercase}
  \defaultfontfeatures[\rmfamily]{Ligatures=TeX,Scale=1}
\fi
\usepackage{lmodern}
\ifPDFTeX\else  
\fi
\IfFileExists{upquote.sty}{\usepackage{upquote}}{}
\IfFileExists{microtype.sty}{% use microtype if available
  \usepackage[]{microtype}
  \UseMicrotypeSet[protrusion]{basicmath} % disable protrusion for tt fonts
}{}
\makeatletter
\@ifundefined{KOMAClassName}{% if non-KOMA class
  \IfFileExists{parskip.sty}{%
    \usepackage{parskip}
  }{% else
    \setlength{\parindent}{0pt}
    \setlength{\parskip}{6pt plus 2pt minus 1pt}}
}{% if KOMA class
  \KOMAoptions{parskip=half}}
\makeatother
\usepackage{xcolor}
\makeatletter
\ifx\paragraph\undefined\else
  \let\oldparagraph\paragraph
  \renewcommand{\paragraph}{
    \@ifstar
      \xxxParagraphStar
      \xxxParagraphNoStar
  }
  \newcommand{\xxxParagraphStar}[1]{\oldparagraph*{#1}\mbox{}}
  \newcommand{\xxxParagraphNoStar}[1]{\oldparagraph{#1}\mbox{}}
\fi
\ifx\subparagraph\undefined\else
  \let\oldsubparagraph\subparagraph
  \renewcommand{\subparagraph}{
    \@ifstar
      \xxxSubParagraphStar
      \xxxSubParagraphNoStar
  }
  \newcommand{\xxxSubParagraphStar}[1]{\oldsubparagraph*{#1}\mbox{}}
  \newcommand{\xxxSubParagraphNoStar}[1]{\oldsubparagraph{#1}\mbox{}}
\fi
\makeatother

\usepackage{longtable,booktabs,array}
\usepackage{calc} % for calculating minipage widths
\usepackage{etoolbox}
\makeatletter
\patchcmd\longtable{\par}{\if@noskipsec\mbox{}\fi\par}{}{}
\makeatother
\IfFileExists{footnotehyper.sty}{\usepackage{footnotehyper}}{\usepackage{footnote}}
\makesavenoteenv{longtable}
\usepackage{graphicx}
\makeatletter
\def\maxwidth{\ifdim\Gin@nat@width>\linewidth\linewidth\else\Gin@nat@width\fi}
\def\maxheight{\ifdim\Gin@nat@height>\textheight\textheight\else\Gin@nat@height\fi}
\makeatother
\setkeys{Gin}{width=\maxwidth,height=\maxheight,keepaspectratio}
\makeatletter
\def\fps@figure{htbp}
\makeatother

\makeatletter
\@ifpackageloaded{caption}{}{\usepackage{caption}}
\AtBeginDocument{%
\ifdefined\contentsname
  \renewcommand*\contentsname{Table of contents}
\else
  \newcommand\contentsname{Table of contents}
\fi
\ifdefined\listfigurename
  \renewcommand*\listfigurename{List of Figures}
\else
  \newcommand\listfigurename{List of Figures}
\fi
\ifdefined\listtablename
  \renewcommand*\listtablename{List of Tables}
\else
  \newcommand\listtablename{List of Tables}
\fi
\ifdefined\figurename
  \renewcommand*\figurename{Figure}
\else
  \newcommand\figurename{Figure}
\fi
\ifdefined\tablename
  \renewcommand*\tablename{Table}
\else
  \newcommand\tablename{Table}
\fi
}
\@ifpackageloaded{float}{}{\usepackage{float}}
\floatstyle{ruled}
\@ifundefined{c@chapter}{\newfloat{codelisting}{h}{lop}}{\newfloat{codelisting}{h}{lop}[chapter]}
\floatname{codelisting}{Listing}

\makeatother
\makeatletter
\@ifpackageloaded{caption}{}{\usepackage{caption}}
\@ifpackageloaded{subcaption}{}{\usepackage{subcaption}}
\makeatother

\ifLuaTeX
  \usepackage{selnolig}  % disable illegal ligatures
\fi
\usepackage[]{natbib}
\usepackage{bookmark}

\IfFileExists{xurl.sty}{\usepackage{xurl}}{} % add URL line breaks if available
\hypersetup{
  pdftitle={Title},
  pdfauthor={Author 1; Author 2},
  pdfkeywords={3 to 6 keywords, that do not appear in the title},
  colorlinks=true,
  linkcolor={blue},
  filecolor={Maroon},
  citecolor={Blue},
  urlcolor={Blue},
  pdfcreator={LaTeX via pandoc}}

\newcommand{\anon}{1}

\usepackage{amssymb} %maths
\usepackage{amsmath} %maths
\usepackage{amsthm}
\usepackage{makecell}
\usepackage{mathrsfs} %free version for \mathscr meaning math script capital letter
\usepackage{listings}
\usepackage{bm}
\usepackage[utf8]{inputenc} %useful to type directly diacritic characters
\usepackage{algorithm, algorithmic}
\usepackage{natbib}
\usepackage{hyperref}
\hypersetup{
    colorlinks=true,
    linkcolor=blue,      % Color of internal links (equations, figures)
    citecolor=blue,      % Color of citations
    urlcolor=cyan        % Color of external URLs
}
\usepackage{graphicx}
\usepackage{multirow}
\usepackage{threeparttable}
\usepackage{listings} %We recommend the listings package for including R code in LaTeX documents.
\usepackage{pdfpages}
\usepackage{enumitem}
\usepackage{booktabs} % For professional-looking tables
\usepackage{textcase} 
\usepackage{setspace}
\newtheorem{theorem}{Theorem}[section]
\newtheorem{corollary}[theorem]{Corollary}
\newtheorem{lemma}[theorem]{Lemma}

\newtheorem{definition}{Definition}[section]

\newcommand{\Matern}{Mat\'ern }

\newcommand{\muone}{\mu_1}
\newcommand{\mumone}{\mu_{-1}}

\newcommand{\Ncalx}{\mathcal{N}} % mu pair for specific x
\newcommand{\Rrisk}{\mathcal{R}} % ITR risk
\DeclareMathOperator*{\infr}{inf} % Infimum operator

\newcommand{\sign}{\operatorname{sign}}

\newcommand{\myT}{T}     % Surrogate loss in ITR
\newcommand{\CT}{C_{\myT}}     % Conditional risk in ITR
\newcommand{\bmx}{\bm{x}}
\newcommand{\bmX}{\bm{X}}
\begin{document}

\def\spacingset#1{\renewcommand{\baselinestretch}%
{#1}\small\normalsize} \spacingset{1}

%%%%%%%%%%%%%%%%%%%%%%%%%%%%%%%%%%%%%%%%%%%%%%%%%%%%%%%%%%%%%%%%%%%%%%%%%%%%%%
\if1\anon
{
    \title{\bf A General Theory of Outcome Weighted Learning for Individualized Treatment Rules}
  \author{Zhu Wang 
      %\thanks{
    %The author gratefully acknowledges \textit{please remember to list all relevant funding sources in the version that gives all author information}
      %}
      \hspace{.2cm}\\
    %Department of Preventive Medicine, 
      University of Tennessee Health Science Center\\
    Memphis, TN, USA\\
   E-mail: zwang145@uthsc.edu
}
\maketitle
} \fi

\if0\anon
{
  \bigskip
  \bigskip
  \bigskip
  \begin{center}
    {\LARGE\bf A General Theory of Outcome Weighted Learning for Individualized Treatment Rules}

\end{center}
  \medskip
} \fi

\bigskip
\begin{abstract}
Personalized medicine addresses conditions with heterogeneous treatment responses by identifying individualized treatment rules (ITRs), a key objective within the field of policy learning. Outcome weighted learning (OWL) is a prominent framework for estimating ITRs that casts the problem as weighted classification, thereby directly targeting clinical benefit while leveraging machine learning tools. Existing theory for OWL, however, has largely been developed for specific surrogate losses and
primarily for Gaussian kernels. In particular, systematic convergence-rate results remain limited for broad classes of machine learning losses and for Matérn kernels. These kernels are particularly relevant as their finite, tunable smoothness often reflects real-world data more accurately than the Gaussian kernel, which they include as a limiting case. In this article, we establish a quantitative relationship between the population 0--1 risk and the population risk induced by a nonnegative surrogate loss via a constrained variational transformation of the loss. This transform admits simplifications for convex losses and yields closed-form expressions for selected nonconvex losses. We show that the resulting relationship provides nontrivial upper bounds on excess 0--1 risk under a positive-definiteness condition on the transform. Building on these bounds, we derive convergence rates for kernel-based OWL with convex and bounded nonconvex losses under either (i) smoothness assumptions paired with \Matern\ kernels or (ii) geometric noise conditions paired with Gaussian kernels. We further investigate robust nonconvex losses, including robust binomial losses, and develop an iteratively reweighted convex optimization algorithm applicable to OWL and residual weighted learning. Numerical studies, including simulations and an application to the AIDS Clinical Trials Group Study 175, illustrate the practical performance of the proposed methods. Supplementary materials are available online.

\end{abstract}

\noindent%

\vfill

\newpage
\spacingset{1.75} % changed to increase line 26!
%\spacingset{1.8} % DON'T change the spacing!

%\makeatletter
%\newcommand\subsubsubsection{\@startsection{paragraph}{4}{\z@}{-2.5ex\@plus -1ex \@minus -.25ex}{1.25ex \@plus .25ex}{\normalfont\normalsize\bfseries}}
%\newcommand\subsubsubsubsection{\@startsection{subparagraph}{5}{\z@}{-2.5ex\@plus -1ex \@minus -.25ex}{1.25ex \@plus .25ex}{\normalfont\normalsize\bfseries}}
%\makeatother

%\usepackage{titlesec}

%\setcounter{secnumdepth}{4}

%\titleformat{\paragraph}
%{\normalfont\normalsize\bfseries}{\theparagraph}{1em}{}
%\titlespacing*{\paragraph}
%{0pt}{3.25ex plus 1ex minus .2ex}{1.5ex plus .2ex}

%\title{Policy Learning, Convexity, Robustness, and Risk Bounds}

%%%blinding
%\begin{document}
%\maketitle
%\tableofcontents

%\author{}
%\date{}

%\begin{document}

%\begin{abstract}

%\end{abstract}
%\noindent {\bf Keywords:} Outcome weighted learning; Convex optimization; Machine learning; Kernel method,  Sobolev space; Convergence rate

\section{Introduction}

Personalized medicine, or precision medicine, marks a shift from traditional ``one-size-fits-all'' healthcare toward tailoring prevention, diagnosis, and treatment strategies based on individual characteristics such as genetics, biomarkers, lifestyle, and environmental factors \citep{kosorok2019precision}. Central to this paradigm are individualized treatment rules (ITRs), decision functions that map patient-specific data to treatments designed to maximize clinical outcomes \citep{tsiatis2019dynamic}. Within the broader field of policy learning, ITRs represent optimal decision policies learned from observational or trial data using statistical learning methods, with the aim of enhancing therapeutic efficacy while minimizing adverse effects.

This article considers a two-arm randomized clinical trial where each patient provides data $(\boldsymbol{X}, A, R)$: $\boldsymbol{X}$ represents $m$-dimensional baseline covariates, $A \in \{1, -1\}$ is the assigned treatment, and $R \geq 0$ denotes the clinical outcome or reward. The objective is to identify an optimal rule $d(\boldsymbol{X})$ that maximizes the expected outcome across the population, quantified by the value function $V(d)$:
\begin{equation}\label{eqn:value}
    V(d) = \mathbb{E}[R \mid A = d(\boldsymbol{X})] = \mathbb{E}\left[\frac{R}{\pi(A, \boldsymbol{X})} \mathbb{I}(A = d(\boldsymbol{X}))\right],
\end{equation}
where $\pi(a, \boldsymbol{x}) := P(A = a \mid \boldsymbol{X} = \boldsymbol{x})$ is the propensity score, fixed by the study design, and $\mathbb{I}(\cdot)$ is the indicator function \citep{qian2011performance, zhao2012estimating}.

Standard approaches for ITR estimation include regression-based frameworks such as $Q$-learning, which identify optimal decisions by modeling patient outcomes. Rooted in reinforcement learning, $Q$-learning estimates the conditional mean of the outcome for specific patient traits and treatments \citep{qian2011performance}. This framework extends naturally to multi-stage decision-making for dynamic treatment regimes, which adapt as a patient's health evolves over time \citep{chakraborty2013statistical, song2015penalized}.

In contrast, outcome weighted learning (OWL) reframes ITR estimation as a weighted classification task rather than a regression problem \citep{zhao2012estimating}. By minimizing the complement of the value function, OWL optimizes treatment assignments by penalizing misclassifications with reward-based weights. This approach utilizes surrogate loss functions to replace the discontinuous 0--1 loss, thereby enhancing computational tractability. OWL has since been extended to residual-weighted learning (RWL), which further refines decision boundaries for more effective interventions \citep{zhou2017residual}. Unlike $Q$-learning, OWL does not require a correctly specified model for the conditional mean of outcomes, providing robustness against model misspecification. Furthermore, the OWL framework is versatile enough to accommodate complex data structures, including survival outcomes and multi-stage regimes \citep{zhao2015new, cui2017tree}.

The No Free Lunch (NFL) theorem in machine learning posits that no single algorithm or loss function can universally outperform all others across every dataset or problem domain \citep{bach2024learning}. This principle motivates the exploration of alternative loss functions within OWL. While margin-based surrogate losses, such as those investigated by \citet{bartlett2006convexity}, are widely utilized for classification, a systematic theoretical foundation for their application to OWL remains underdeveloped.

In OWL as in standard classification problems, three error sources arise: estimation error due to finite samples, approximation error from a restricted function class $\mathcal{H}$ that an estimator comes from, and approximation error resulting from the use of a surrogate loss in place of the 0--1 loss. The pivotal step to analyze the last source is to establish a quantitative relationship between the surrogate risk and the 0--1 risk. \citet{bartlett2006convexity} introduced classification
calibration and the $\psi$-transform, which yield sharp bounds on the excess misclassification risk in terms of the excess surrogate risk. Related extensions to other settings were discussed by \citet{Stei:2007}. 
Within OWL, analogous bounds have been obtained for the hinge loss and the smoothed ramp loss \citep{zhao2012estimating, zhou2017residual}. \citet{jiang2019entropy}, along with the accompanying commentaries, established excess-risk bounds for the entropy loss.
In these settings, the raw lower bound on the conditional excess risk ($\tilde{\Psi}$) is convex, allowing Jensen’s inequality to bound the global expected risk. However, for general loss functions, $\tilde{\Psi}$ may lack convexity. Furthermore, while the general OWL framework proposed by \citet{huang2019multicategory} ensures asymptotic risk consistency, its theoretical error bound remains computationally intractable. Because calculating this bound requires identifying the global extrema of inverse-propensity-weighted outcomes across the entire covariate space, their result serves primarily as a structural existence proof rather than a practically computable bound.

In contrast to this prior work, our article extends \citet{bartlett2006convexity} to develop a practically computable excess risk bound for general loss functions, which can be used directly to establish convergence rates for OWL. To address the lack of convexity in general settings, we utilize the biconjugate transform $\Psi$ to convert the raw $\tilde\Psi$ into a convex function, enabling the application of Jensen's inequality to bound the global expected risk. We demonstrate that for any
measurable decision function, the relationship between the corresponding excess risks is characterized by the $\Psi$-transform. We show that the $\Psi$-transform is a generalization of the $\psi$-transform, thus sharing many common properties 
while also exhibiting distinct features unique to the OWL setting. We derive explicit forms of the $\Psi$-transform for commonly used loss functions. 
This $\Psi$-transform often simplifies to a form analogous to the $\psi$-transform, consistent with the results established by \citet{zhao2012estimating, zhou2017residual, jiang2019entropy}. 
Regarding the latter work on the entropy or binomial loss, the assumptions made directly on the outcomes are stronger than the constraints placed on the conditional expectations in the $\Psi$-transform. This highlights how the $\Psi$-transform serves as a direct generalization of the $\psi$-transform beyond bounded conditional probabilities.
Furthermore, we apply the $\Psi$-transform to the
robust nonconvex loss functions recently developed by \citet{wang2024unified}. 

After establishing the risk relationship, we analyze the excess surrogate risk, which decomposes into
estimation error from kernel methods due to finite samples and the approximation error arising from the restricted function class $\mathcal{H}$. In addition to the Gaussian kernel used in \citet{zhao2012estimating, zhou2017residual}, we also employ Matérn kernels, which offer tunable smoothness through a shape parameter and recover the Gaussian kernel as a limiting case \citep{porcu2024matern, bach2024learning}. This choice leverages the flexibility of reproducing kernel Hilbert spaces (RKHSs) to produce stable estimators with tractable theoretical guarantees. Using concentration inequalities \citep{bach2024learning}, we derive nonasymptotic bounds on the estimation error for both convex and nonconvex loss functions. We then control the approximation error under two complementary sets of assumptions. First, we assume that the target function belongs to a Sobolev space that is norm-equivalent to the Matérn RKHS. This connection quantifies regularity through derivative integrability and suggests that Matérn-based procedures can adapt to unknown smoothness. Under this framework, we obtain rates of order \(n^{-1/2}\) for convex losses and \(n^{-1/3}\) for nonconvex losses. Second, we impose geometric noise conditions on the decision boundary, without requiring explicit smoothness assumptions on the underlying distribution \citep{steinwart2007fast, zhao2012estimating, zhou2017residual}. Under these conditions, we recover the same rates for general convex and bounded nonconvex loss functions, thereby extending prior results for specific losses to a broader class of learners.

The contributions of this article are summarized as follows. 
First, we introduce the notion of policy calibration, identifying it as a pointwise form of Fisher consistency for OWL, and showing that it is equivalent to classification calibration. 
Second, we develop the \(\Psi\)-transform for OWL, which yields upper bounds on the excess 0--1 risk in terms of the excess surrogate risk. We characterize conditions under which these bounds are nontrivial and under which surrogate loss-risk consistency implies Bayes-risk consistency for OWL. We also derive simplified expressions for the \(\Psi\)-transform for a range of convex and selected nonconvex loss functions. 
Third, we establish convergence rates for kernel-based OWL under two complementary sets of assumptions: smoothness conditions for \Matern\ kernels and geometric noise conditions for Gaussian kernels. Our results cover both convex and bounded nonconvex loss functions. 
Finally, we extend robust OWL and RWL to concave--convex (CC) surrogate losses \citep{wang2024unified} and propose an iteratively reweighted convex optimization (IRCO) algorithm within the majorization--minimization framework. 
Convex analysis techniques are used extensively throughout the article, spanning from the theoretical methods to the algorithmic implementation.

The remainder of the article is organized as follows. Section~\ref{sec:no2} studies the \(\Psi\)-transform and develops a general framework relating excess 0--1 risk to excess surrogate risk in OWL; it also shows how convexity, or specific forms of nonconvexity, simplify the computation of \(\Psi\). 
Section~\ref{sec:no3} presents convergence-rate results for convex and nonconvex losses under the \Matern\ smoothness assumption and the geometric noise assumption for Gaussian kernels. 
Section~\ref{sec:est} describes kernel-based OWL with smooth convex losses and with smooth nonconvex losses via the IRCO algorithm; both approaches are further adapted to incorporate residuals. 
Section~\ref{sec:sim} evaluates the proposed methods through simulations and presents an application to the AIDS Clinical Trials Group Study 175 \citep{hammer1996trial} in Section~\ref{sec:data}. 
Section~\ref{sec:dis} concludes with a discussion. All proofs are provided in the supplementary material.

\section{Relating Excess Risk to Excess Surrogate Risk}\label{sec:no2}
Analyzing the approximation error of surrogate losses over the 0--1 loss, this section assumes $\mathcal{H}$ contains all measurable functions and focuses on population expectations.
%\subsection{Notation}
We define the true 0--1 risk as
\(
\mathcal{R}(f)=\mathbb{E}\left[\frac{R}{\pi(A, \bm{X})} \mathbb{I}(A \neq \operatorname{sign}(f(\bm{X})))\right],
\)
and the Bayes risk as $\mathcal{R}^{*}=\inf _{f}\{\mathcal{R}(f) \mid f: \mathcal{X} \rightarrow \mathbb{R}\}$, where $f^*$ is a minimizer such that 
$\mathcal{R}^{*}=\mathcal{R}\left(f^{*}\right)$.  The optimal ITR is given by $d^{*}(x)=\operatorname{sign}\left(f^{*}(x)\right)$. 
The excess risk is $\mathcal{R}(f) - \mathcal{R}^*$. 
Similarly, the $T$-risk for nonnegative surrogate loss $T$ is
\(
\mathcal{R}_T(f)=\mathbb{E}\left[\frac{R}{\pi(A, \bm{X})} T(A f(\bm{X}))\right],
\)
 and its minimal value is
$\mathcal{R}_{T}^{*}=\inf _{f}\{\mathcal{R}_{T}(f) \mid f : \mathcal{X} \rightarrow \mathbb{R}\}$. The excess surrogate risk is $\mathcal{R}_T(f) - \mathcal{R}_{T}^{*}$.
Denote the conditional mean outcomes 
$\mu_a(\bmx) = \mathbb{E}[R | \bm{X}=\bmx, A=a]$. 
The optimal decision is given by $d^*(\bm{x}) = \operatorname{sign}(\mu_1(\bm{x}) - \mu_{-1}(\bm{x}))$.
Define the conditional $T$-risk,
\[
\mathbb{E}\left[\frac{R}{\pi(A, \bm{X})} T(A f(\bm{X}))\mid \bmX = \bmx\right] = 
    \mu_1(\bm{x}) T(f(\bmx)) + \mu_{-1}(\bm{x}) T(-f(\bmx)).
\]
%It is helpful to view the conditional $T$‑risk in terms of 
Consider generic conditional mean outcomes $\bm\mu=(\muone, \mumone)$, a generic classifier value $p \in \mathbb{R}$, and the generic conditional $T$-risk,
\[
    C_T(p, \bm\mu) = \mu_1 T(p) + \mu_{-1} T(-p).
\]
The generic conditional $T$-risk matches the conditional $T$‑risk of $f$  at a given $\bmx \in \mathcal{X}$ when we substitute $\muone=\muone(\bmx), \mumone=\mumone(\bmx)$ and $p = f(\bmx)$. Here, variations in $p$ 
represent variations in the value of $f$ at that fixed $\bmx$. 
Define the optimal conditional $T$-risk ($C_T^*$) and the optimal conditional $T$-risk for the wrong sign ($C_T^-$) as follows:
\begin{align*}
    C_T^*(\bm\mu) &= \infr_p C_T(p, \bm\mu),\\
    C_T^-(\bm\mu) &= \infr_{p : p \cdot \sign(\mu_1-\mu_{-1}) \le 0} C_T(p, \bm\mu).
\end{align*}

\subsection{The $\psi$-transform for OWL}
Since OWL is a weighted classification, we begin with a minimal
condition that can be viewed as a pointwise form of Fisher consistency
for classification \citep{bartlett2006convexity}.
\begin{definition}[Classification-Calibration]
    A loss function $T$ is classification-calibrated if, for all $\eta \in [0, 1]$ with $\eta \neq 1/2$, 
    \(
    H_T^{-}(\eta)>H_T^*(\eta),
    \)
where $C_T(\eta) = \eta T(p) + (1-\eta)T(-p), H^*_T(\eta) = \inf_{p \in \mathbb{R}} C_T(\eta),
H^{-}_T(\eta) = \inf_{p : p(2\eta-1) \le 0} C_T(\eta).$
\end{definition}
Define $\psi$ as the Fenchel-Legendre biconjugate of $\tilde{\psi}(\theta) = H^-_T(\frac{1+\theta}{2}) - H^*_T(\frac{1+\theta}{2})$. The following theorem establishes that selected $\psi$-transforms, originally developed for classification on the domain $[-1, 1]$ but extended to $[0, \infty)$, can be applied to OWL to bound excess risks.
\begin{theorem}\label{thm:n23}
    Assume $T$ is classification-calibrated and $\psi$ is positive homogeneous, i.e., for $c > 0$ and $\theta \geq 0$, the equality holds:
   \(
   \psi(c\theta) = c \psi(\theta).
   \)
    Then  
    \(   \psi(\mathcal{R}(f) - \mathcal{R}^*) \leq \mathcal{R}_T-\mathcal{R}_T^*\) and $\psi(\theta) = \theta \psi(1)$ for all $\theta \in [0,\infty)$.
\end{theorem}

This generalizes earlier results to bound the excess risk for OWL using the hinge loss \citep{zhao2012estimating} and the smoothed ramp loss \citep{zhou2017residual}, for which $\psi(\theta) = \theta$. However, many other loss functions do not satisfy the conditions of Theorem~\ref{thm:n23} (see Table~\ref{tab:pc1}). To address this limitation, we
introduce a new framework for OWL.
\subsection{Policy-Calibration}
In classification-based ITR estimation, a method is Fisher consistent if its optimal decision function $f^*$ induces the reward-maximizing treatment rule \citep{zhao2012estimating}, ensuring that minimizing the surrogate risk validly proxies minimizing the true risk. While this property has been established for specific loss functions in OWL \citep{zhao2012estimating, zhou2017residual, jiang2019entropy} and generalized by \citet{huang2019multicategory, zhang2020multicategory}, our work takes a different approach. The following condition provides a constructive, pointwise form of this consistency, extending classification-calibration \citep{bartlett2006convexity}. 
\begin{definition}[Policy-Calibration]\label{def:n6}
     A loss function $T$ is policy-calibrated if, for all
     $\muone \neq \mumone$,
     \(
     C_T^-(\bm{\mu}) > C_T^*(\bm{\mu}).
     \)
 \end{definition}
 Minimizing the $T$-risk is equivalent to pointwise minimization of the conditional $T$-risk. Policy-calibration requires that incorrect signs     yield strictly higher risk than the optimum.
 As shown in the theorem below, the two types of calibration are equivalent. 
 \begin{theorem}\label{thm:no5}
     Loss function $T$ is policy-calibrated if and only if $T$ is classification-calibrated. Loss function $T$ is Fisher consistent for OWL if     and only if $T$ is Fisher consistent for  classification.
 \end{theorem}

This result is stronger for binary treatments than \citet[Theorem 1]{huang2019multicategory}, which established only the one-way implication that classification calibration implies policy calibration. 
Moreover, it immediately yields Fisher consistency criteria for both general loss functions \citep{lin2004note} and truncated loss functions \citep{wu2007robust} in OWL. Related criteria can also be found in earlier work, including \citet{huang2019multicategory} and \citet{zhang2020multicategory}. In particular, Corollary~\ref{lem:itrfisher} has been established in these prior works.
\begin{corollary}\label{lem:itrfisher}
    Assume a loss function $T$ satisfies:
    \begin{align*}
        T(p) &< T(-p) \quad \forall p > 0, \quad T'(0) \neq 0.
    \end{align*}
Then $T$ is Fisher consistent for OWL. % it is more complex to prove policy-calibration 
\end{corollary}

\begin{corollary}\label{cor:trunfisher}
Let $\ell$ be a nonincreasing function with $\ell'(0) < 0$. For a fixed threshold $t$, the truncated loss function is defined as $T(p) = \min\{\ell(p), \ell(t)\}$. Then, for any $t \leq 0$, $T$ is Fisher consistent for OWL. % it is more complex to prove policy-calibration
\end{corollary}

\subsection{The $\Psi$-Transform and the Relationship Between Excess Risks}
We first define a functional $\Psi$-transform of the loss function. We then heuristically relate the $\Psi$-transform to the $\psi$-transform. Theorem~\ref{thm:n1} subsequently establishes that the $\Psi$-transform provides upper bounds on excess risk through the excess $T$-risk.
\begin{definition}[$\Psi$-Transform]\label{def:n7}
    Let \(\bm{\mu} = (\mu_{1}, \mu_{-1}), M=\sup_{\bm\mu}(\muone + \mumone) < \infty\).
    For $v \in [0, M]$,
    define
    \begin{align*}
        \Ncalx(v)
&= \{(\muone, \mumone): \muone \geq 0, \mumone \geq 0, |\muone - \mumone| = v, \muone + \mumone \leq M\}.
    \end{align*}
    %Assume that 
    %    \(C_T^-(\bm{\mu}) - C_T^*(\bm{\mu})\) is defined for all $\bm{\mu} \in \mathcal{N}(v)$, and define its infimum
    \begin{equation*}
        \label{eqn:tPsi}
        \tilde{\Psi}(v) = \inf_{\bm{\mu} \in \Ncalx(v)} C_T^-(\bm{\mu}) - C_T^*(\bm{\mu}).
    \end{equation*} 
    Furthermore, define the function $\Psi:[0, M] \to [0, \infty)$ by 
    \( \Psi = (\tilde{\Psi})^{**} \), the Fenchel-Legendre biconjugate of $\tilde{\Psi}$.
\end{definition}
The nonnegativity of $\Psi$ follows from Lemma~\ref{lem:no2}, part 7. We illustrate the relationship between  
$\tilde\Psi$ and $\tilde\psi$. Assume 
$\muone \geq \mumone, M = \infty$. We obtain
\[\tilde{\Psi}(\muone - \mumone)=\inf_{\muone-\mumone=v}C_T^{-}(\bm{\mu})-C^*_T(\bm{\mu}).\]
After a change of variables and adding a trivial constraint, $\tilde\psi$ can be rewritten as 
\[\tilde{\psi}(\eta-(1-\eta))=\inf_{\eta-(1-\eta)=2\eta-1}H_T^{-}(\eta)-H_T^*(\eta).\]
Setting $\muone = \eta, \mumone=1-\eta$ and $v = 2\eta-1$ yields the same minimization operation in symbolic form. 
%However, a limitation is that \(C_T^{-}(\bm{\mu}) - C_T^{*}(\bm{\mu})\) is defined only for \(\bm{\mu} \in \mathcal{N}(v)\). This restriction is needed to establish part~1 of the following theorem, which in turn underscores the central role of the \(\Psi\)-transform in deriving risk bounds.
The following theorem demonstrates the fundamental role of the $\Psi$-transform.  
\begin{theorem}\label{thm:n1}
    \leavevmode
    \begin{enumerate}
        \item[1.] For any nonnegative loss function $T$, any measurable $f: \mathcal{X} \rightarrow \mathbb{R}$, and any probability distribution on $\mathcal{X}\times\mathcal{A}\times\mathfrak{R}$, we have:
            \( \Psi\left(\mathcal{R}(f)-\mathcal{R}^{*}\right) \leq \mathcal{R}_T(f)-\mathcal{R}_T^{*}. \)
        \item[2.] 
Assume that $\Psi$ is positive definite, meaning
    $\Psi(0) = 0, \Psi(v) > 0$ for all $v > 0$.
Then:
\begin{enumerate}
    \item[a.] For any sequence $\left(\theta_{i}\right)$ in $[0,M]$,
%$$
\(\Psi\left(\theta_{i}\right) \rightarrow 0 \text { if and only if } \theta_{i} \rightarrow 0.\)
%$$

\item[b.] For every sequence of measurable functions $f_{i}$ : $\mathcal{X} \rightarrow \mathbb{R}$ and every probability distribution on $\mathcal{X}\times\mathcal{A}\times\mathfrak{R}$,
%$$
\(\mathcal{R}_T\left(f_{i}\right) \rightarrow \mathcal{R}_T^{*} \text { implies that }  \mathcal{R}\left(f_{i}\right) \rightarrow \mathcal{R}^{*}.\)
%$$

\item[c.] $T$ is policy-calibrated.
\end{enumerate}
    \end{enumerate}
\end{theorem}

This shows that policy calibration is weaker than positive definiteness. Although the $\Psi$‑transform is positive semidefinite, meaning $\Psi(0) = 0, \Psi(v) \geq 0$ for $v > 0$ by Lemma~\ref{lem:no2}, parts 7 and 8, it may still fail to be invertible on $[0, M]$ because $\Psi$ need not be positive definite, even when $T$ is policy‑calibrated. Nevertheless, common loss functions yield a positive definite $\Psi$‑transform, as demonstrated in Table~\ref{tab:pc1}. Consequently, by Lemma~\ref{lem:no2}, part 9, we can validly express the upper bound on excess risk.

For a convex $T$, the following theorem generalizes \citet[Theorem 2]{bartlett2006convexity} by establishing that policy-calibration is equivalent to a derivative condition at the origin, yielding a stronger result than Corollary~\ref{lem:itrfisher}.
This result has been established by \citet{huang2019multicategory} and is listed here for completeness. Furthermore,
convexity also yields a simplified $\Psi$-transform similar to the $\psi$-transform.
\begin{theorem}\label{thm:compPsiconvex}
    Suppose $T$ is a nonnegative loss function. 
    \begin{enumerate}
        \item If $T$ is convex, then $T$ is policy-calibrated if and only if $T$ is differentiable at 0 and $T'(0)<0$.
        \item 
            If $T$ is convex and policy-calibrated, then
            \[{\Psi}(v) = \inf_{\bm{\mu} \in \Ncalx(v)}\left[C_T(0, \mathcal{M}_x) - C_T^*(\mathcal{M}_x)\right] =
                \inf_{S \in [v, M]}\left[S\cdot T(0) - C_T^*\left(\frac{S+v}{2}, \frac{S-v}{2}\right)\right]
            . \]
    \end{enumerate}
\end{theorem}

This is used in Table~\ref{tab:pc1} (with details provided in the supplementary material) to compute the $\Psi$-transform for loss functions commonly used in machine learning. The $\psi$-transform for most of these loss functions was computed by \citet{bartlett2006convexity}. While the smoothed ramp loss is nonconvex, its computation is simplified by Lemma~\ref{lem:no2}, part 10. 
\begin{table}[h!]
    \centering
    \begin{tabular}{lccc}
        \toprule
%        \textbf{Loss}       & \textbf{Function}              & \textbf{\shortstack{Classification\\Calibration}} & \textbf{\shortstack{Policy\\Calibration}} \\
        \textbf{Loss}       & \textbf{Function}              & \textbf{$\psi$-Transform} & \textbf{$\Psi$-Transform} \\
        \midrule
        \midrule
        Expo         & $\exp(-p)$                    & $1-\sqrt{1-v^{2}}$                                                  & $M\left(1-\sqrt{1-(\frac{v}{M})^2}\right)$ \\
        \hline
        TQ %quadratic
                     & $\left(\max(1-p, 0)\right)^2$ & $v^2$                                                  & $\frac{v^2}{M}$ \\
                     \hline
        Hinge               & $\max (1-p, 0)$               & $v$                                                  & $v$ \\
        \hline
        Distance
        %\shortstack{Distance-weighted\\ discrimination} 
                            & $\begin{cases} \frac{1}{p}                        & \text{if } p \ge \gamma \\ 
                                \frac{1}{\gamma}\left(2-\frac{p}{\gamma}\right) & \text{if } p < \gamma   \\ 
                            \end{cases} $
                            & $\frac{v}{\gamma}$ & $\frac{v}{\gamma}$ \\
                            & $\gamma > 0$                  &             &             \\
                            \hline
        ARC-X4 & $|1-p|^k, k > 1$ & 
        \makecell{
            \(1 - 2^{k-1} \left(1 - v^2\right)\times\) \\[10pt]
            \(\left[(1-v)^{\frac{1}{k-1}} + (1+v)^{\frac{1}{k-1}}\right]^{1-k}\)
        }
               & \makecell{
                   \(M - 2^{k-1}(M^2-v^2)\times\) \\[10pt]
                   \(\left[(M-v)^{\frac{1}{k-1}} + (M+v)^{\frac{1}{k-         1}}\right]^{1-k}\)
               }\\
               %& $M - \frac{2^{k-1}(M^2-v^2)}{\left[(M+v)^{\frac{1}{k-1}} + (M-v)^{\frac{1}{k-         1}}\right]^{k-1}}$ \\
               %ARC-X4 & $|1-u|^k, k > 1$ & $1-\frac{2^{k-1}\left(1-v^{2}\right)}{\left[(1-v)^{\frac{1}{k-1}}+(1+v)^{\frac{1}{k-1}}\right]^{k-1}}$ & $M - \frac{2^{k-1}(M^2-v^2)}{\left[(M+v)^{\frac{1}{k-1}} + (M-v)^{\frac{1}{k-         1}}\right]^{k-1}}$ \\
               \hline
               Sigmoid & $1 -  \tanh(kp), k >0$ & $v$ & $v$ \\
               \hline
               Binomial & $\log(1+\exp(-p))$ & \makecell[l]{
                   \(\frac{1+v}{2} \log(1+v)\) \\[\jot] % Use \\[<length>] to add space. \jot is a standard small math space.
                   \(+ \frac{1-v}{2} \log(1-v)\)
               }
& \makecell[l]{
    \, \(\frac{M+v}{2}\log(M+v)\) \\[\jot]
    \(+ \frac{M-v}{2}\log(M-v)\) \\[\jot]
    \(-M \log M\)
}\\
\hline
SRamp & $
\begin{cases}0 & p \geq 1 \\ (1-p)^{2} & 0 \leq p < 1 \\ 2-(1+p)^{2} & -1 \leq p < 0 \\ 2 & p<-1\end{cases}
$           & $v$    & $v$\\
\bottomrule
\end{tabular}
\caption{$\psi(v)$ for $v \in [0, 1]$ and $\Psi(v)$ for $v \in [0, M]$, where $M < \infty$ can be relaxed for certain loss functions. Expo: exponential, TQ: truncated quadratic, Distance: Distance-weighted discrimination, SRamp: smoothed ramp.}
\label{tab:pc1}
\end{table}

For the hinge, distance‑weighted discrimination, sigmoid, and smoothed ramp loss functions, the $\Psi$‑transform extends the $\psi$‑transform because they share the same function values, with domains $[0,\infty)$ and $[0,1]$, respectively. For other loss functions, the constraint $\mu_1 + \mu_{-1} \le M < \infty$ is needed for computational validity, causing $\Psi$ and $\psi$ to differ; rescaling rewards so that $M=1$ makes them coincide.
The $\Psi$‑transform also yields a risk bound equivalent to that of \citet{zhao2012estimating} for the hinge loss and to that of \citet{zhou2017residual} for the smoothed ramp loss. 

In the remainder of this section, we present two lemmas analogous to the classification-specific results  
in \citet{bartlett2006convexity}, both of which are essential to the proof of Theorem~\ref{thm:n1}. 

\begin{lemma}\label{lem:n2_4}
    For the excess risk of a decision function $f$, we have the identity
    \[
        \mathcal{R}(f) - \mathcal{R}^* = \mathbb{E}_{\bm{X}}\Big[\mathbb{I}(\text{sign}(f(\bm{X})) \neq d^*(\bm{X})) \cdot 
        |\mu_1(\bm{X}) - \mu_{-1}(\bm{X})|\Big].
    \]
\end{lemma}
\begin{lemma}\label{lem:no2}
    For any nonnegative loss function $T$, the functions $C_T^*$, $C_T^-$, and $\Psi$ have the following properties:
    \begin{enumerate}
        \item $\CT^*$ and $\CT^-$ are symmetric about $(\muone, \mumone)$: 
            $C_T^*(\mu_1, \mu_{-1}) = C_T^*(\mu_{-1}, \mu_1), C_T^-(\mu_1, \mu_{-1}) = C_T^-(\mu_{-1}, \mu_1)$.
        \item 
            $C_T^*(\mu_1, \mu_{-1})$ is concave on $(\mu_1, \mu_{-1})$, and 
            \(\CT^*(\muone, \mumone) \leq \CT^*\left(\frac{\muone + \mumone}{2}, \frac{\muone + \mumone}{2}\right)= \CT^-\left(\frac{\muone + \mumone}{2}, \frac{\muone + \mumone}{2}\right).\)

        \item 
            If $T$ is policy-calibrated and 
            $\mu_1 \ne \mu_{-1}$, then 
            \(\CT^*(\muone, \mumone) < \CT^*\left(\frac{\muone + \mumone}{2}, \frac{\muone + \mumone}{2}\right).\)

        \item 
            $C_T^{-}$ is concave on $\{ (\mu_1, \mu_{-1}) \in \mathbb{R}^2 \mid \mu_1 \leq \mu_{-1} \}$ and on $\{ (\mu_1, \mu_{-1}) \in \mathbb{R}^2      \mid \mu_1 \geq \mu_{-1} \}$, and 
            $C_T^-(\mu_1, \mu_{-1}) \ge C_T^*(\mu_1, \mu_{-1})$.

        \item If $T$ is continuous, then $C_T^*$ and $C_T^-$ are continuous in $(\mu_1, \mu_{-1})$.

        \item 

            $\Psi$ is continuous. $\tilde{\Psi}$ is continuous if $\mu_1 + \mu_{-1}$ is bounded or $\CT^- - \CT^*$ is coercive.

        \item $\Psi$ is nonnegative and minimal at 0.

        \item \textbf{$\Psi(0)=0$}.

        \item If $\Psi$ is positive definite on $[0, M]$, then $\Psi$ is strictly increasing and invertible on $[0, M]$.
        
        \item $\Psi = \tilde\Psi$ if and only if $\tilde\Psi$ is convex.
    \end{enumerate}
\end{lemma}
Lemma~\ref{lem:no2} parallels Lemma~2 in \citet{bartlett2006convexity}. Except for part~9, $\tilde\Psi$ and $\Psi$ share many properties with their counterparts $\tilde\psi$ and $\psi$. However, proving Lemma~\ref{lem:no2} requires more intricate arguments. For instance, continuity of $\Psi$ in part 6 relies  on the Berge Maximum Theorem \citep{aliprantis2006infinite} and inverse image results 
\citep{rudin1976principles}. 

\subsection{The $\Psi$-Transform of Nonconvex Loss Functions}

Computing the $\Psi$‑transform for nonconvex loss functions generally relies on computing $\tilde\Psi$, and directly evaluating the biconjugate is nontrivial unless $\tilde\Psi$ is convex. However, the following theorem provides a simplified approach.

\begin{theorem}
    \label{thm:crit}
Denote $T(\infty)=\lim_{p \to \infty}T(p), T(-\infty)=\lim_{p\to -\infty}T(p)$.
    Suppose function $T$ satisfies the following assumptions:
    \begin{enumerate}[label=A\arabic*:]
        \item $T$ is nonincreasing, bounded and $T(\infty) = 0$.
        \item 
            $T(p)+T(-p) \geq T(\infty)+T(-\infty) \forall p\in\mathbb{R}$.
        \item $T(0) \geq T(-\infty)/2$, 
        \item The constrained infimum occurs at the boundaries:
            \begin{align*}
                \inf_{p \leq 0} C(p) &= \min\left(C(0), \lim_{p \to -\infty} C(p)\right) &\text{ if } \mu_1 > \mu_{-1},\\
                \inf_{p \geq 0} C(p) &= \min\left(C(0), \lim_{p \to \infty} C(p)\right)  &\text{ if } \mu_1 < \mu_{-1}.
            \end{align*}
    \end{enumerate}
    Then, for all $v\geq0, \Psi(v) = vT(0).$ 
\end{theorem}

Condition A1 is a standard assumption satisfied by many robust loss functions \citep{wu2007robust, park2011robust, wang2024unified}. 
Establishing A2 can be simplified through two sufficient conditions that are often more straightforward to verify.
\begin{lemma}\label{lem:n8} 
    Assume that $T$ is differentiable for all $p\in\mathbb{R}$ and $ T'(p) \le T'(-p) \text{ for all } p > 0$.
    Then condition A2 of Theorem~\ref{thm:crit} holds.
\end{lemma}

\begin{figure}[htpb!]
        \centering
        \includegraphics[width=0.4\textheight]{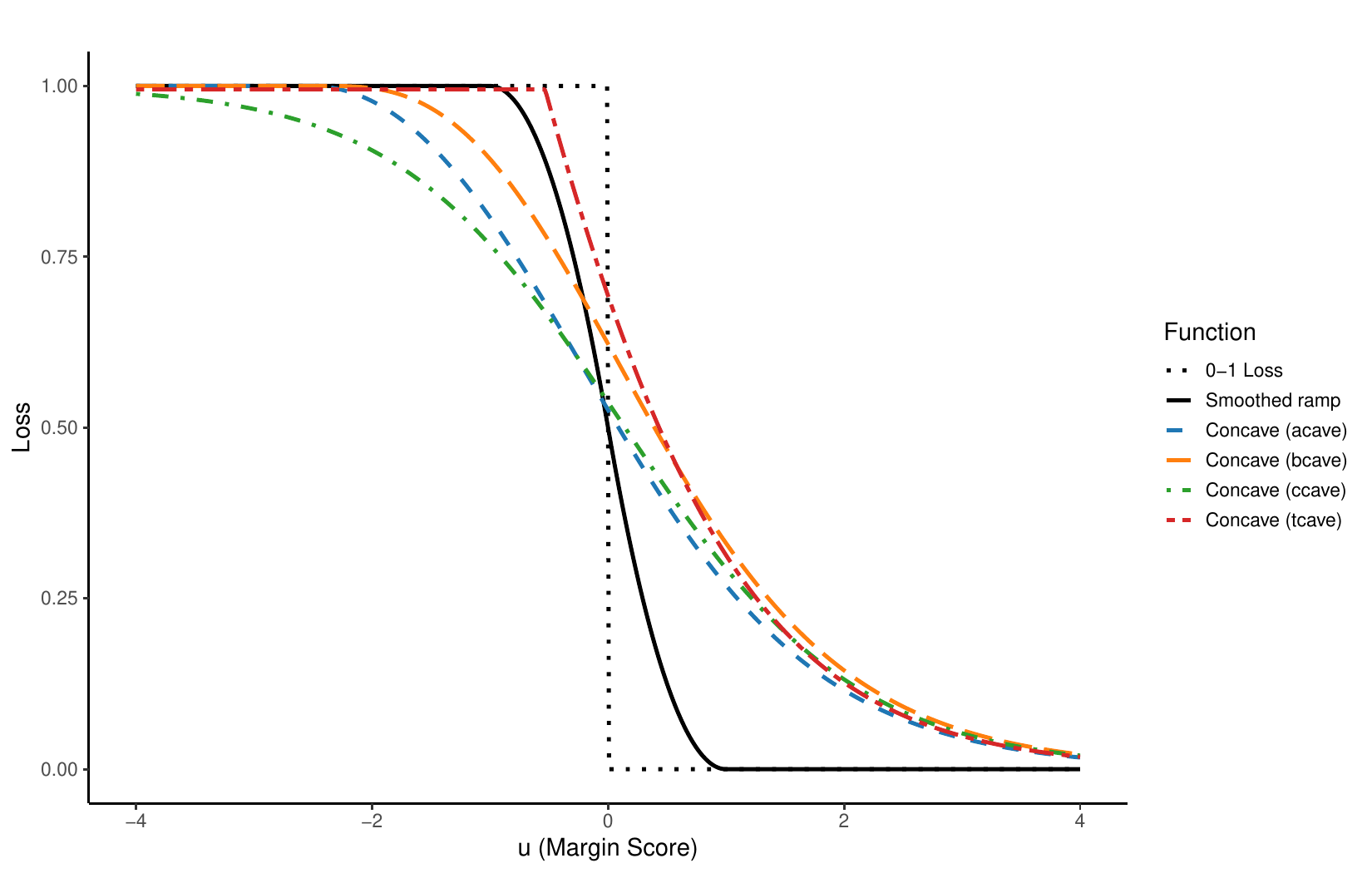}
        \caption{Loss function comparisons: $g\circ s$, where $s$ denotes the binomial loss. Various  concave functions $g$ are shown, with their associated $\sigma$ selected to closely approximate the scaled smoothed ramp loss.}
        \label{fig:loss_comp}
    \end{figure}

\begin{lemma}\label{lem:main_n10} 
    Assume $g$ is nondecreasing and concave, $s$ is nonnegative convex, $T=g \circ s$, and
    \(
    g(0)+g(2s(0)) \geq T(\infty)+T(-\infty)
    \).
    Then condition A2 of Theorem~\ref{thm:crit} holds.
\end{lemma}
We focus on the robust binomial loss functions $T = g \circ s$ within the CC‑family \citep{wang2024unified}, where $s$ is the convex binomial component and $g$ is the concave component specified in Table~\ref{tab:concave_sigma}. The resulting robust nonconvex loss functions are illustrated in Figure~\ref{fig:loss_comp}. The following theorem summarizes  their key properties when used as surrogate loss functions.
\begin{theorem}\label{thm:cc}
    Let the loss function be defined as $T=g\circ s$, where $s(p)=\log(1+\exp(-p))$, $g$ and its range of robustness parameter $\sigma^2$ are given in Table~\ref{tab:concave_sigma}. Then $T$ is policy-calibrated, its $\Psi$-transform is $\Psi(v) = vT(0)$ for $v \geq 0$, and $T$ is Lipschitz continuous.
\end{theorem}

\begin{table}[h]
    \centering
    \begin{tabular}{cccc}
        \hline
        Concave & Function $g$ & Range of $\sigma^2$ & $C$\\
        \hline
        acave &$\begin{cases} \frac{1}{2}(1-\cos((2z)^{1/2}/\sigma) & \text{if } z \leq \sigma^2\pi^2/2\\
            1,                                   & \text{otherwise}
        \end{cases}$
              &$\left(\frac{2\log 2}{\pi^2 }, \frac{4 \log 2}{\pi^2 }\right)$   &$\frac{1}{2\sigma^2}$\\ 
            bcave &$1-(1-2z/\sigma^2)^3     I(z \le \sigma^2/2)$ &$\left(2\log2, \frac{2 \log 2}{1 - 2^{-1/3}}\right]$ & $\frac{6}{\sigma^2}$\\
            ccave &$1 - \exp{(-z/\sigma^2)}$ &$(0, 1)$ &$\frac{1}{\sigma^2}$\\
            tcave &$\min(\sigma, z)$ &$[(\log 2)^2, (2\log 2)^2]$ &1\\
            \hline
        \end{tabular}
        \caption{Concave function $g$, range of robustness parameter $\sigma^2$, and 
    Lipschitz constant $C$.}
        \label{tab:concave_sigma}
    \end{table}
    \section{Learning Rates}\label{sec:no3}
Building on the risk bounds in Section~\ref{sec:no2}, we now focus our analysis on the excess surrogate risk. Assume observations $\left\{\left(\boldsymbol{x}_i, a_i, r_i\right)\right\}_{i = 1}^n$ are independent and identically distributed random variables. 
Let $\hat{f}_\lambda$ be a minimizer of:
 $$
 \inf_{f \in \mathcal{H}} \left(
 \frac{1}{n}             \sum_{i=1}^{n} \frac{r_{i}}{\pi\left(a_{i}, \boldsymbol{x}_{i}\right)} T\left(a_{i}           f\left(\boldsymbol{x}_{i}\right)\right)
  +   \frac{\lambda}{2}\|f\|_\mathcal{H}^{2}
\right),
 $$
  where $\lambda > 0$ is a regularization parameter and $\mathcal{H}$ is an RKHS associated with a Mercer kernel $k$. We assume the kernel
  is bounded, $\sup_{\boldsymbol{x} \in \mathcal{X}} \sqrt{k(\boldsymbol{x}, \boldsymbol{x})} = \mathcal{K} < \infty$, which holds for Matérn, Gaussian, and linear kernels on compact domains \citep{hastie2009elements, bach2024learning}.
Analysis of the excess surrogate risk typically begins with decomposing it into estimation and approximation error. The estimation error reflects stochastic fluctuations from finite sampling, while the approximation error is a deterministic quantity depending on $\mathcal{H}$ and the underlying data distribution. We first examine the approximation error, as it permits a unified analysis that is independent of the specific loss function employed. The estimation error is addressed in the subsequent section, as its analysis depends on the specific loss type (e.g., convex or non-convex).
    \subsection{Approximation Error Assumptions}

    The approximation error is defined as: 
    \begin{equation}\label{eqn:appr0}
        \rho(b) = \inf_{f \in \mathcal{H}}\left(
            \mathcal{R}_{T}({f}) - \mathcal{R}_{T}^{*}
        + \frac{b}{2} \|{f}\|_\mathcal{H}^2\right), 
    \end{equation}
    where $b > 0$ is a constant depending on the loss function $T$. We analyze this error using
    smoothness and geometric noise assumptions.

    \subsubsection{Smoothness Assumption}
The Matérn kernel is defined as:
$$k(\boldsymbol{x}, \boldsymbol{x}') = \frac{2^{1-\alpha}}{\Gamma(\alpha)} \left( \sqrt{2\alpha} \frac{d}{\varrho} \right)^\alpha K_\alpha \left( \sqrt{2\alpha} \frac{d}{\varrho} \right),$$
where $d = \|\boldsymbol{x} - \boldsymbol{x}'\|$, $\varrho$ is the bandwidth, $\alpha$ is the smoothness parameter, $\Gamma$ is the Gamma function, 
and $K_\alpha$ is the modified Bessel function of the second kind.
This kernel spans a range of functional smoothness: it yields the exponential kernel $(\alpha = 0.5)$ for rough or noisy data, the Matérn 3/2 kernel $(\alpha = 1.5)$ for once-differentiable processes, and converges to the Gaussian kernel $(\alpha \to \infty)$ for extremely smooth functions. The resulting RKHS is equivalent to a Sobolev space of order $\alpha + m/2$ 
\citep{porcu2024matern, bach2024learning}. The Sobolev space of order $t > 0$, denoted by $H^t(\mathbb{R}^m)$, consists of functions $f \in L^2(\mathbb{R}^m)$ satisfying the following condition in the Fourier domain:
\begin{equation*}\label{eqn:sobolev}
\int_{\mathbb{R}^m} (1 + \|\omega\|_2^2 )^t |\hat{f}(\omega)|^2 \, d\omega \leq C_s < \infty,
\end{equation*}
where $\hat{f}$ denotes the Fourier transform of $f$. Assume the target function $f^* \in H^t(\mathbb{R}^m)$.

It can be shown that (\ref{eqn:appr0}) yields a minimization problem for      some $\nu > 0$:
%\begin{equation}\label{eqn:main_n75}
 \(   Q(\nu, f^*)=\inf_{f\in\mathcal{H}} \left\{ \|f-f^*\|_{L_2(P)} +\nu\|f\|^2_{\mathcal{H}} \right\}\), which 
%\end{equation}
%The optimization problem (\ref{eqn:main_n75}) 
can be conveniently carried out in the      Fourier domain. 
     We begin with the Radon-Nikodym theorem to upper bound the $L_2(P)$-norm by the           $L_2(\mathbb{R}^m)$-norm.
Assuming $P$ has a bounded density $\|p\|_{\infty}$, we upper bound $Q(\nu, f^*)$ by $\tilde{Q}(\nu, f^*)$ using the $L_2(\mathbb{R}^m)$-norm:
\begin{equation*}\label{eqn:main_n72}
    \tilde{Q}(\nu, f^*) = \inf_{f\in \mathcal{H}} \left\{ \|p\|_\infty \|f-f^*\|^2_{L_2(\mathbb{R}^m)} + \nu \|f\|^2_\mathcal{H} \right\}.
\end{equation*}
Since the Mat\'{e}rn kernel is translation-invariant, its RKHS norm and $L_2$ distance can be expressed in the Fourier domain via Bochner’s and Parseval’s theorems. 
For the Mat\'ern kernel, the Fourier transform is given by:
     \begin{equation*}\label{eqn:matern_fourier}
         \hat{q}(\omega) = C_k (1 + \varrho^2 \|\omega\|_2^2)^{-\alpha},
     \end{equation*} 
     where $C_k$ is a constant of proportionality. This formulation leads to the RKHS norm: 
%The Mat\'{e}rn Fourier transform $\hat{q}(\omega) \propto (1 + \varrho^2 \|\omega\|_2^2)^{-\alpha}$ leads to the RKHS norm:
%\begin{equation*}\label{eqn:n77}
 \(   \|f\|_{\mathcal{H}}^{2} = \frac{1}{(2 \pi)^{m}} \int_{\mathbb{R}^{m}} \frac{|\hat{f}(\omega)|^{2}}{\hat{q}(\omega)} d \omega.\)
%\end{equation*}
Combining this with the Fourier representation of $\|f-f^*\|_{L_2(\mathbb{R}^m)}^2$ yields a direct frequency-domain minimization, summarized in the following lemma.
    \begin{lemma}\label{lem:matern}
        Assume $T$ is a $C$-Lipschtiz continuous function, $\frac{R}{\pi(A, \bm{X})}\leq B$ almost surely, $f^* \in H^t(\mathbb{R}^m)$, $\mathcal{H}$ is an RKHS associated with a \Matern kernel of order $\alpha$, and $b > 0$, 
        then we have:
        \begin{equation*}\label{eqn:appr}
            \inf_{f \in \mathcal{H}} \left( \mathcal{R}_T(f) - \mathcal{R}_T^* + \frac{b}{2} \|f\|_\mathcal{H} \right)
            \leq 
            \begin{cases}
                \frac{b}{\sqrt{2}}\|f^*\|_\mathcal{H}  &\text{ if } t \geq \alpha\\
                b^{t/\alpha}
                2^{1/2-t/\alpha}
                (BC)^{1-t/\alpha}
                \sqrt{
                    \frac{
                C_s \|p\|_\infty^{1-t/\alpha} }{(2\pi)^m C_k^{t/\alpha}}} &\text{ if } t < \alpha.
            \end{cases}
        \end{equation*}

            \end{lemma}

            \subsubsection{Geometric Noise Assumption}
Geometric noise quantifies the difficulty of resolving the decision boundary $\Delta = \{\bmx : \mu_1(\bmx) = \mu_{-1}(\bmx)\}$. When $|\mu_1(\bmx) - \mu_{-1}(\bmx)|$ concentrates near $\Delta$, the boundary is harder to approximate. This is captured by the geometric noise exponent $q > 0$, where larger $q$ indicates a smoother boundary. For an RKHS $\mathcal{H}$ with a Gaussian kernel $k(\bmx, \bmx') = \exp(-\|\bmx - \bmx'\|^2 / 2\varrho^2)$, the approximation error depends on the bandwidth $\varrho$ and exponent $q$. By setting $\varrho(\lambda) = \lambda^{-1/((q+1)m)}$, there exists $c > 0$ such that for all $\lambda > 0$:
\begin{equation}\label{eqn:geom}
    \rho(\lambda) = \inf_{f \in \mathcal{H}} \left( \mathcal{R}_T(f) - \mathcal{R}_T^* + \frac{\lambda}{2} \|f\|_\mathcal{H}^2 \right) \leq c \lambda^{q /(q+1)}.
\end{equation}
See \citet[Lemma 3.9]{zhou2017residual} and \citet[Theorem 2.7]{steinwart2007fast} for details.
            \subsection{Learning Rates for Convex Loss Functions}

            Decompose the excess surrogate risk as:
            \begin{equation}\label{eqn:main_no50}
                \begin{aligned}
                    \mathcal{R}_{T}(\hat{f}_\lambda) - \mathcal{R}_{T}^{*} &\le \left[ \frac{\lambda}{2}                         \|\hat{f}_\lambda\|_\mathcal{H}^2 +       \mathcal{R}_{T}(\hat{f}_\lambda) - \inf_{f \in                             \mathcal{H}}\left(\frac{\lambda}{2} \|f\|_\mathcal{H}^2 +               \mathcal{R}_{T}(f)\right) \right] \\
                                                                           &\quad + \inf_{f \in                           \mathcal{H}}\left(\frac{\lambda}{2}            \|f\|_\mathcal{H}^2 +           \mathcal{R}_{T}(f) -                  \mathcal{R}_{T}^{*}\right).
                \end{aligned}
            \end{equation}
            The following theorem bounds the estimation error contained in the bracketed term in \eqref{eqn:main_no50}. 

            \begin{theorem}\label{thm:n10}
                Assume that the convex loss function $T$ is $C$-Lipschitz-continuous and $\frac{R}{\pi(A, \bm{X})} \leq B$ almost surely. Then
                \begin{align}\label{eqn:thm42} 
                    \mathbb{E}[\mathcal{R}_T(\hat{f}_{\lambda})] -                 \mathcal{R}_T(f^*)
                    \leqslant 
                    \frac{24 B^2C^{2}\mathcal{K}^{2}}{\lambda  n} + \inf_{f \in \mathcal{H}}\left(\mathcal{R}_T(f) - \mathcal{R}_T(f^*)+    \frac{\lambda}{2}\|f\|_{\mathcal{H}}^{2}\right).
                \end{align} 
            \end{theorem}
            The proof in the supplementary material leverages strong convexity and concentration inequalities, and converts high-probability tail bounds into an expectation inequality. By adapting the approach from \citet[Proposition 4.6]{bach2024learning} for OWL, we provide a direct argument that avoids a proof by contradiction.
            Characterizing the approximation error through functional smoothness, Lemma~\ref{lem:matern} and Theorem~\ref{thm:n10} yield the following convergence rates for Matérn kernels under convex surrogate loss functions.
                \begin{corollary}\label{cor43}
                    Assume that the conditions stated in Theorem~\ref{thm:n10} are satisfied, $f^*\in H^t(\mathbb{R}^m)$, and $\mathcal{H}$ is the RKHS associated with a \Matern kernel of order $\alpha$. 
                    For any $f \in \mathcal{H}$, 
                    setting the regularization parameter $\lambda(f) = \frac{4\sqrt{3}BC\mathcal{K}}{\sqrt{n}\|{f}\|_\mathcal{H}}$
                    minimizes the upper bound in \eqref{eqn:thm42}, establishing the following bounds:
                    \begin{align*}
                        \mathbb{E}[\mathcal{R}_T(\hat{f}_\lambda)] - \mathcal{R}_T^* 
                        &\leq
                                                                                         \begin{cases}
                                                                                             n^{-\frac{1}{2}}4\sqrt{6}BC\mathcal{K}
                                                                                             \|f^*\|_\mathcal{H}  &\text{ if } t \geq \alpha\\
                                                                                             %\frac{4\sqrt{6}BC\mathcal{K}}{\sqrt{n}}\|f^*\|_\mathcal{H}  &\text{ if } t \geq \alpha\\
                                                                                             n^{-\frac{t}{2\alpha}}
                                                                                             BC
                                                                                             \sqrt{2}(4\sqrt{3}\mathcal{K})^{t/\alpha}
                                                                                             \sqrt{\frac{
                                                                                             C_s \|p\|_\infty^{1-t/\alpha} }{(2\pi)^m C_k^{t/\alpha}}} &\text{ if } t < \alpha.
                                                                                         \end{cases}
                                                                                         \end{align*}
                    \end{corollary}
In the best-case scenario where $f^* \in \mathcal{H}$ ($t \geq \alpha$), the approximation error is negligible, yielding a dimension-independent rate of $O(n^{-1/2})$ with $\lambda = \frac{4\sqrt{3}BC\mathcal{K}}{\sqrt{n}\|f^*\|_{\mathcal{H}}}$. 
If only first-order derivatives are assumed to be square-integrable ($t=1$), for the exponential kernel with $\alpha = (m+1)/2$, the rate is $O(n^{-1/(m+1)})$. 
For intermediate smoothness $t \in [1, \alpha]$, the rate $O(n^{-t/(2\alpha)})$, or $O(n^{-t/(m+1)})$, interpolates between these extremes. This result highlights the adaptive nature of kernel methods: 
for a fixed kernel of order $\alpha$, the excess surrogate risk scales according to the target's true smoothness $t$. This adaption occurs automatically, obviating the need for prior knowledge of the regularity of the underlying function space.

When the smoothness of $f^*$ is unknown, we rely on the geometric noise assumption to establish the convergence rate for the Gaussian kernel.
                    \begin{corollary}\label{cor:convex2}

                        Let $\mathcal{X}$ be the closed unit ball of the Euclidean space $\mathbb{R}^{m}$, and $P$ be a distribution on $\mathcal{X} \times \mathcal{A} \times \mathcal{M}$ that has geometric noise exponent $0<q<\infty$ with constant $c$ given in (\ref{eqn:geom}). 
                        Consider the RKHS     $\mathcal{H}$ generated by a Gaussian kernel with bandwidth $\varrho(\lambda)=\lambda^{\frac{1}{(q+1)m}}$, where $\lambda=n^{-\frac{q+1}{2q+1}}$.
                        Suppose the conditions stated in Theorem~\ref{thm:n10} are satisfied.
                        Then 
                        we have
                        \begin{align*}
                            \mathbb{E}[\mathcal{R}_T(\hat{f}_\lambda)] - \mathcal{R}_T^* 
&\leq \left(24B^2 C^2 \mathcal{K}^2 +c\right) n^{-\frac{q}{2q+1}}.
                        \end{align*}
                    \end{corollary}
                    Corollary~\ref{cor:convex2} establishes a learning rate of \( O\left(n^{-\frac{q}{2q+1}}\right) \), which approaches $n^{-1/2}$ as the geometric noise exponent $q$ increases.
                    While \citet{zhao2012estimating} provides a similar result for the hinge loss, our corollary generalizes this to all Lipschitz continuous convex loss functions.
                    \subsection{Learning Rates for Nonconvex Loss Functions}
                    Decompose the excess surrogate risk as:
                    \begin{equation}\label{eqn:main_deco3}
                        \begin{aligned}
                            \mathcal{R}_{T}(\hat{f}_\lambda) - \mathcal{R}_{T}^{*} 
           & \le \left[
               \left( \Rrisk_{T, n}(\breve{f}_\lambda)
               - \Rrisk_T(\breve{f}_\lambda) \right)
               + \left(\Rrisk_T(\hat{f}_\lambda) - \Rrisk_{T, n}(\hat{f}_\lambda)\right) 
           \right]  \\
           &\quad +  
           \inf_{f \in                           \mathcal{H}}\left(\frac{\lambda}{2}            \|f\|_\mathcal{H}^2 +           \mathcal{R}_{T}(f) -                  \mathcal{R}_{T}^{*}\right), 
                        \end{aligned}
                    \end{equation}
                    where $\breve{f}_{\lambda}$ is a minimizer of (\ref{eqn:appr0}).
                    The following theorem bounds the estimation error contained in the bracketed term in \eqref{eqn:main_deco3} by applying concentration inequalities and adapting the proof in \citet{zhou2017residual} to general bounded nonconvex loss functions.
                    \begin{theorem}\label{thm:n45}
                        Assume that the loss function $T$ is bounded such that $0 \leq T(p) \leq D$, $C$-Lipschitz-continuous, and $\frac{R}{\pi(A, \boldsymbol{X})} \leq B$ almost surely. Then
                        \begin{equation}\label{eqn:thm45}
                            \mathbb{E}[\mathcal{R}_T(\hat{f}_\lambda)] - \mathcal{R}_T^{*} 
                            \leq 
                            BD \sqrt{\frac{\pi}{2n}} +
                            2BC \mathcal{K} \sqrt{\frac{2 BD}{\lambda n}}+ 
                            \inf_{f \in \mathcal{H}} \left(
                            \mathcal{R}_T(f) - \mathcal{R}_T^* + \frac{\lambda}{2} \|f\|_\mathcal{H}^2 \right).
                        \end{equation}

                    \end{theorem}
Combining the smooth approximation error assumption, Theorem~\ref{thm:n45}, 
                    Lemma~\ref{lem:matern}, and Young's inequality for products \citep[Lemma 7.1]{steinwart2008support}, we obtain the learning rates for \Matern kernels under nonconvex loss functions.
                    \begin{corollary}\label{cor:n46}
                        Assume that the conditions of Theorem~\ref{thm:n45} are satisfied, 
$f^*       \in H^t(\mathbb{R}^m)$, and $\mathcal{H}$ is the RKHS associated with a \Matern kernel of order $\alpha$. 
                        For any $f \in \mathcal{H}$, setting the regularization parameter 
                        \(
                        \lambda(f) = \left( \frac{C_1}{\|{f}\|_{\mathcal{H}}^2} \right)^{2/3},
                        \)
                        where $C_1 = \frac{2BC}{\sqrt{n}} \mathcal{K} \sqrt{2 BD}$, 
                    minimizes the upper bound in \eqref{eqn:thm45}, yielding the following bounds:
                        \begin{align*}
                            \mathbb{E}[\mathcal{R}_T(\hat{f}_\lambda)]-\mathcal{R}_T^{*} 
     &\leq BD\left(\frac{\pi}{2n}\right)^{1/2} +  
     B\left(\frac{\mathcal{K}^2C^2D}{n}\right)^{1/3} \\ 
     &\quad+ \begin{cases}
         B\left(\frac{\mathcal{K}^2C^2D}{n}\right)^{1/3}
         2^{5/6} \|f^*\|_\mathcal{H} &\text{ if } t \geq \alpha\\
         \left(\frac{\mathcal{K}^2 D}{n}\right)^{t/(3\alpha)} (BC)^{1 - t/(3\alpha)}
         2^{1/2 + t/\alpha} 
         \sqrt{\frac{
         C_s \|p\|_\infty^{1-t/\alpha} }{(2\pi)^m C_k^{t/\alpha}}} &\text{ if } t < \alpha.
     \end{cases}
     \end{align*}   

                        \end{corollary}
                        Compared to convex loss functions, the well-specified case ($f^* \in \mathcal{H}, t \geq \alpha$) achieves a dimension-independent rate of $O(n^{-1/3})$ by setting $\lambda = (C_1 / \|f^*\|_{\mathcal{H}}^2)^{2/3}$. For the exponential kernel ($\alpha = (m+1)/2$), $t=1$ results in rate of $O(n^{-2/(3m+3)})$. At intermediate smoothness ($t \in [1, \alpha]$), the rate $O(n^{-2t/(3\alpha)})$, or $O(n^{-2t/(3m+3)})$, spans these extremes. This result again underscores the
                        adaptive nature of kernel methods, as the estimator automatically achieves the optimal rate for the target function's regularity $t$.
                        
Without explicit smoothness assumptions, we rely on the geometric noise condition to establish the convergence rate for the Gaussian kernel.
                        \begin{corollary}\label{cor:noncon}
                            Let $\mathcal{X}$ be the closed unit ball of the Euclidean space $\mathbb{R}^{m}$, and $P$ be a distribution on $\mathcal{X} \times \mathcal{A} \times \mathcal{M}$ that has geometric noise exponent $0<q<\infty$ with constant $c$ given in \eqref{eqn:geom}. 
                            Consider the RKHS     $\mathcal{H}$ generated by a Gaussian kernel with bandwidth $\varrho(\lambda)=\lambda^{\frac{1}{(q+1)m}}$, where $\lambda=n^{-\frac{q+1}{3q+1}}$. 
                            Assume that the conditions of Theorem~\ref{thm:n45} are satisfied.
                            Then we have
                            \[
                                \mathbb{E}[\mathcal{R}_T(\hat{f}_\lambda)-\mathcal{R}_T^{*}] 
                                \leq \left(2BC\mathcal{K}\sqrt{2BD} + c\right)n^{-\frac{q}{3q+1}}.
                            \]
                        \end{corollary}
Corollary~\ref{cor:noncon} establishes a learning rate of $O(n^{-q/(3q+1)})$, which approaches $n^{-1/3}$ as $q$ increases. While \citet{zhou2017residual} presents a comparable result, our derivation extends to the broader class of general bounded nonconvex loss functions.

    \section{Computational Implementation}\label{sec:est}
    We implement kernel methods for OWL and RWL using both convex and nonconvex smooth loss functions. 
    \subsection{Algorithms for OWL}\label{sec:posrewards}
   We aim to solve the following optimization problem:
\begin{equation}\label{eqn:n25}
    \min_{f \in \mathcal{H}, \delta \in \mathbb{R}} \quad 
    \frac{1}{n} \sum_{i=1}^{n} \frac{r_i}{\pi(a_i, \bm{x}_i)}T(a_i (f(\boldsymbol{x}_i) + \delta))
    +\frac{\lambda}{2}\|f\|_\mathcal{H}^{2}, 
\end{equation}
where $\delta$ is a bias term \citep{steinwart2008support}. By the representer theorem \citep{kimeldorf1971some}, the solution $f(\boldsymbol{x})$ can be expressed as a finite linear combination of kernels: $f(\boldsymbol{x}) = \sum_{j=1}^{n} v_{j} k(\boldsymbol{x}, \boldsymbol{x}_{j})$. This transforms \eqref{eqn:n25} into a finite-dimensional optimization over $(\boldsymbol{v}, \delta)$:
\begin{equation}\label{eqn:emloss1}
    \min_{\boldsymbol{v} \in \mathbb{R}^n, \delta \in \mathbb{R}} \quad 
    \frac{1}{n} \sum_{i=1}^n \frac{r_i}{\pi(a_i, \bm{x}_i)}T\left(a_i\left(\sum_{j=1}^n v_j k(\boldsymbol{x}_i, \boldsymbol{x}_j) + \delta\right)\right)
    %+\frac{\lambda}{2} \boldsymbol{v}^\top \mathbf{K} \boldsymbol{v},
    +\frac{\lambda}{2} \sum_{i,j=1}^n v_i v_j k(\boldsymbol{x}_i, \boldsymbol{x}_j).
\end{equation}
%where $\mathbf{K}$ is the kernel matrix with entries $K_{ij} = k(\boldsymbol{x}_i, \boldsymbol{x}_j)$. 
%Unlike the hinge loss, the binomial loss and its robust variants typically result in non-sparse solutions where most coefficients $v_i$ are non-zero. 

For convex $T$, we solve \eqref{eqn:emloss1} using the limited-memory Broyden–Fletcher–Goldfarb–Shanno (L-BFGS) algorithm \citep{nocedal2006numerical}. 
    For nonconvex $T=g\circ s$ in the CC-family, 
we propose an IRCO algorithm. Denote the gradient of $g$ as $g'$ if $g$ is differentiable; otherwise, let $g'$ represent a supergradient. The updated weights at the $(j+1)$-th iteration, denoted by $w_i^{(j+1)}$ for $i=1, ..., n$, are computed as follows:
\begin{equation}\label{eqn:no11}
    z_i^{(j+1)}=s(a_i (f_i^{(j)} + \delta^{(j)})), \quad w_i^{(j+1)}=g'(z_i^{(j+1)}).
\end{equation}
The $(j+1)$-th weighted convex subproblem is given by            
    \begin{equation}\label{eqn:emloss2}
        \min_{\boldsymbol{v} \in \mathbb{R}^n, \delta \in \mathbb{R}} \quad 
        \frac{1}{n} \sum_{i=1}^n \frac{w_i^{(j+1)} r_i}{\pi(a_i, \bm{x}_i)} s\left(a_i\left(\sum_{j=1}^n v_j k(\boldsymbol{x}_i, \boldsymbol{x}_j) + \delta\right)\right)
        +\frac{\lambda}{2} \sum_{i,j=1}^n v_i v_j k(\boldsymbol{x}_i, \boldsymbol{x}_j),
    \end{equation}
    which can be solved via the L-BFGS algorithm.
    The algorithm is initialized using the solution to \eqref{eqn:emloss1} with $T=s$.

    \begin{algorithm}[!htbp]
        \caption{IRCO-OWL}
        \label{alg:n1}
        \begin{algorithmic}[1]
            \STATE \textbf{Initialize} $\bm{v}^{(0)}=(v_j^{(1)}, ..., v_j^{(n)})^\top, \delta^{(0)}$ and set $j=0$
            \REPEAT
            \STATE Update weights $w_i^{(j+1)}, i=1, ..., n$ by \eqref{eqn:no11}
            \STATE Update $(\bm{v}^{(j+1)}, \delta^{(j+1)})$ by solving (\ref{eqn:emloss2})
            \STATE $j \leftarrow j + 1 $
            \UNTIL convergence of $(\bm{v}^{(j)}, \delta^{(j)})$
        \end{algorithmic}
    \end{algorithm}

    \subsection{Algorithms for RWL}\label{sec:negrewards}
    RWL reduces outcome variability and ensures balanced treatment assignment; it is also location-scale invariant, meaning estimated rule remains unchanged by shifts or positive scaling of the outcome.
    In RWL, we first model the rewards $r_i$ and replace them with residuals $\hat{r}_i$. Applying this strategy to the residual magnitudes results in a modified loss function \citep{zhou2017residual, chen2018estimating}:
    \begin{equation}\label{eqn:emloss3}
        \begin{split}
        \min_{f \in \mathcal{H}, \delta \in \mathbb{R}} \quad 
        \frac{1}{n} \sum_{i=1}^{n} \frac{|\hat{r}_i|}{\pi(a_i, \bm{x}_i)}
        &\big[I(\hat{r}_i \geq 0)T\big(a_i \big(f(\boldsymbol{x}_i) + \delta\big)\big)\\
        &+ I(\hat{r}_i < 0)T\big( - a_i \big(f(\boldsymbol{x}_i) + \delta\big)\big)\big]
        +\frac{\lambda}{2}\|f\|_\mathcal{H}^{2}, 
        \end{split}
    \end{equation}
    where residuals $\hat{r}_i$ are obtained by minimizing the weighted least squares objective for continuous rewards:
\begin{equation} 
    \min_{\beta_0, \boldsymbol{\beta}} \sum_{i=1}^n \frac{1}{2\pi(a_i, \boldsymbol{x}_i)} (r_i - \beta_0 - \boldsymbol{x}_i^\top \boldsymbol{\beta})^2. 
\end{equation}
    Applying the representer theorem, \eqref{eqn:emloss3} becomes the residual-based optimization problem:
    \begin{equation}\label{eqn:emloss4}
        \begin{split}
        \min_{\boldsymbol{v} \in \mathbb{R}^n, \delta \in \mathbb{R}} \quad 
        \frac{1}{n} \sum_{i=1}^n \frac{|\hat{r}_i|}{\pi(a_i, \bm{x}_i)}
        &\bigg[ I(\hat{r}_i \geq 0) T\bigg(a_i\big(\sum_{j=1}^n v_j k(\boldsymbol{x}_i, \boldsymbol{x}_j) + \delta\big)\bigg)\\
        &+ I(\hat{r}_i < 0) T\bigg(-a_i\big(\sum_{j=1}^n v_j k(\boldsymbol{x}_i, \boldsymbol{x}_j) + \delta\big)\bigg)\bigg]
            +\frac{\lambda}{2} \sum_{i,j=1}^n v_i v_j k(\boldsymbol{x}_i, \boldsymbol{x}_j).
        \end{split}
        \end{equation}
        For convex $T$, we employ the L-BFGS algorithm to solve the weighted optimization problem \eqref{eqn:emloss4}. For nonconvex $T=g\circ s$ in the CC-family, we propose Algorithm~\ref{alg:n2}, which is also based on the IRCO framework. Denote $t_i=1$ if $\hat{r}_i \geq 0$ and $t_i=-1$ otherwise, for $i=1, ..., n$.
The weights at iteration $j+1, w_i^{(j+1)}$ for $i=1, ..., n$, are updated as follows:
\begin{equation}\label{eqn:no16}
            z_i^{(j+1)}=s(t_i a_i (f_i^{(j)} + \delta^{(j)})), \quad w_i^{(j+1)}=g'(z_i^{(j+1)}).
\end{equation}
The $(j+1)$-th weighted convex subproblem is given by            
\begin{equation}\label{eqn:emloss5}
        \begin{split}
        \min_{\boldsymbol{v} \in \mathbb{R}^n, \delta \in \mathbb{R}} \quad 
        \frac{1}{n} \sum_{i=1}^n \frac{|\hat{r}_i|}{\pi(a_i, \bm{x}_i)}
        &\bigg[ I(\hat{r}_i \geq 0) w_i^{(j+1)} s\bigg(a_i\big(\sum_{j=1}^n v_j k(\boldsymbol{x}_i, \boldsymbol{x}_j) + \delta\big)\bigg)\\
        &+ I(\hat{r}_i < 0) w_i^{(j+1)} s\bigg(-a_i\big(\sum_{j=1}^n v_j k(\boldsymbol{x}_i, \boldsymbol{x}_j) + \delta\big)\bigg)\bigg]
            +\frac{\lambda}{2} \sum_{i,j=1}^n v_i v_j k(\boldsymbol{x}_i, \boldsymbol{x}_j),
        \end{split}
    \end{equation}
    which can be solved via the L-BFGS algorithm.
    The algorithm is initialized using the solution to \eqref{eqn:emloss4} with $T=s$.
        \begin{algorithm}[!htbp]
        \caption{IRCO-RWL}
        \label{alg:n2}
        \begin{algorithmic}[1]
            \STATE \textbf{Initialize} $\bm{v}^{(0)}=(v_j^{(1)}, ..., v_j^{(n)})^\top, \delta^{(0)}$ and set $j=0$
            \REPEAT
        %\STATE Compute                   $t_i=1$ if $\hat{r}_i \geq 0$, and $t_i=-1$ otherwise, $i=1, ..., n$
        \STATE Update weights $w_i^{(j+1)}, i=1, ..., n$ by \eqref{eqn:no16}
            \STATE Update $(\bm{v}^{(j+1)}, \delta^{(j+1)})$ by solving (\ref{eqn:emloss5})
            \STATE $j\leftarrow j+1$
            \UNTIL convergence of $(\bm{v}^{(j)}, \delta^{(j)})$
        \end{algorithmic}
    \end{algorithm}
    \section{Simulations}\label{sec:sim}
                        We first conduct simulations of kernel-based OWL using the binomial loss and its robust variants to demonstrate their adaptivity 
                        to varying functional smoothness; subsequently, we compare residual-based OWL methods 
                        against current state-of-the-art approaches.
                        \subsection{Adaptivity of Kernels to Target Smoothness}
                        \noindent Example 1:
                        We generate a univariate covariate $x$ following a uniform distribution on $[-1, 1]$, 
                        and treatment $A \in \{-1,1\}$ with $P(A=1)=0.5$. The response $R$ is log-normal distributed with $\ln(R) \sim N(\tau(x) + \xi(x)a, 1)$, ensuring $R > 0$. We set $\tau(x)=x$ and consider both smooth ($\xi(x) = \sin(4 \pi x)$) and nonsmooth ($\xi(x) = \text{sign}(\sin(4 \pi x))$) target functions. The optimal decision depends on $2\xi(x)$, which we treat as the target function for estimation.
                        
\begin{figure}[htpb!]
                            \centering
                            \includegraphics[width=1\textwidth]{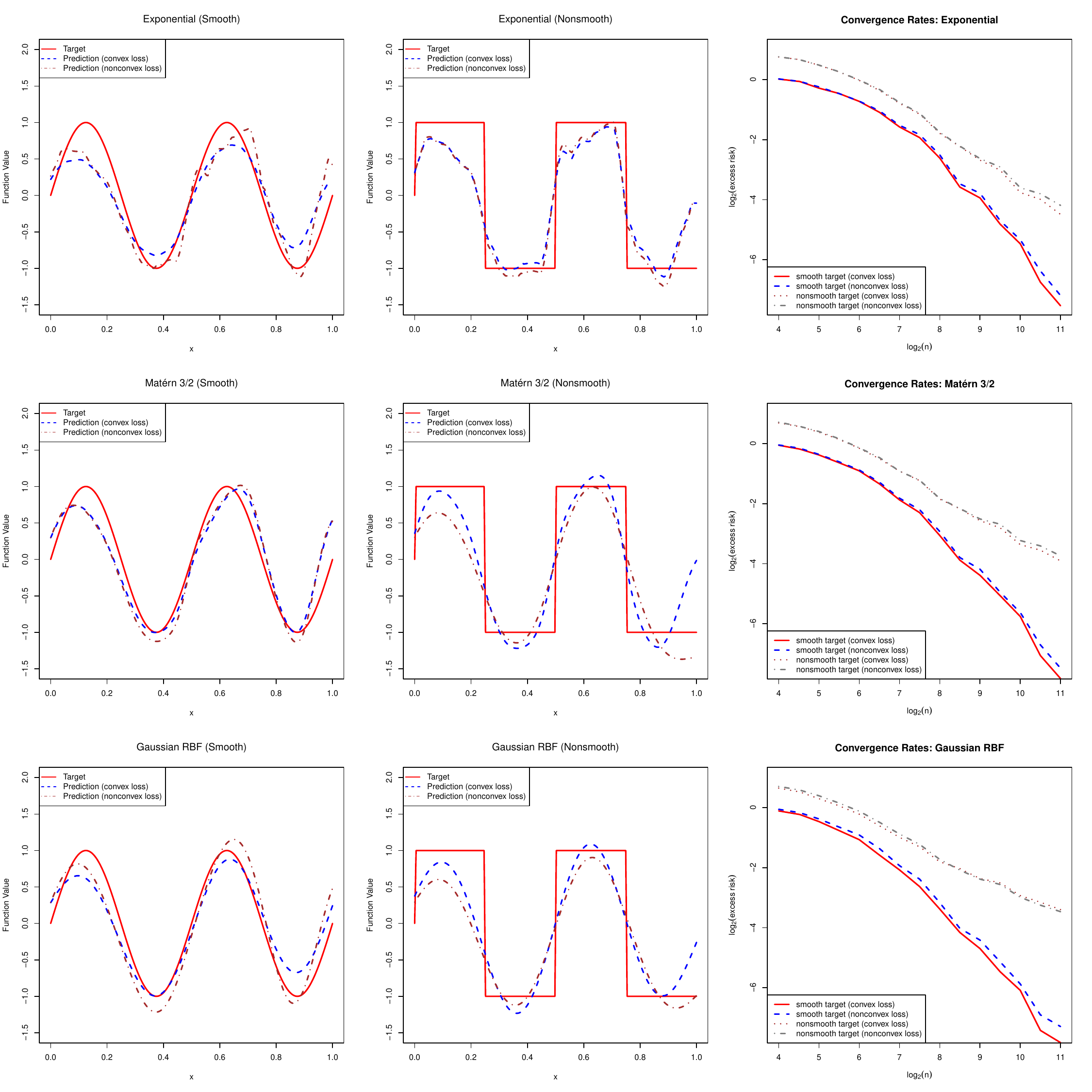} % from high performance computing on UTK ISAAC-NG
                        \caption{Example 1: Comparison of three kernel specifications—exponential, \Matern 3/2, and Gaussian—under both convex and nonconvex loss functions. Left panels: scaled target and estimate. Right panels: excess risks (log-scale) for the smooth and nonsmooth targets.}
                            \label{fig:comp}
                        \end{figure}

                       We evaluate OWL from Section~\ref{sec:posrewards} using binomial (convex) and robust acave-binomial (nonconvex, $\sigma=1$) losses. Using exponential, Matérn 3/2, and Gaussian kernels, we tune hyperparameters to minimize expected risk via $n_\text{test}=10,000$ samples.       Figure~\ref{fig:comp} presents these results: the two left panels display $\xi(x)$ alongside estimates from a single instance ($n=256$), scaled by $1/2$ for comparison. The right panels show convergence rates averaged over 100 replications.
                       See \citet{bach2024learning} for a similar experiment in kernel regression.

                       The results confirm that these kernel methods are adaptive to the underlying regularity, achieving faster convergence rates when the target function is smooth. While the convex loss generally outperforms the nonconvex alternative, the latter performs slightly better under the Gaussian kernel at larger sample sizes. Consistent with kernel regression in \citet{bach2024learning}, kernels with smaller feature spaces (Matérn and Gaussian) underperform on nonsmooth targets relative to the larger feature space of the exponential kernel.
                    \subsection{Comparative Analysis}

                    We conduct a simulation study to investigate the finite-sample performance of the residual-based framework in Section~\ref{sec:negrewards}.
                    We consider both low-dimensional ($m=5$) and moderate-dimensional ($m=50$) settings, where the covariates $x_j$ are sampled independently from a uniform distribution on $[-1, 1]$ for $j=1, ..., m$. The treatment $A \in \{-1, 1\}$ is assigned with $P(A=1)=0.5$, and the response follows $R \sim N(\tau(\bmx) + \xi(\bmx)a, 1)$, where $\tau(\bmx)$ and $\xi(\bmx)$ denote baseline and interaction effects, respectively. We evaluate four scenarios across smooth and nonsmooth functions, with Example 4 adapted from \citet{zhou2017residual}.

                        \noindent Example 2. $\tau(x) = 1 + x_1 + x_2 + 2x_3 + 0.5x_4,
                        \xi(x) = 0.146 + \sin(4  \pi  x_1) + x_2^2$.

                        \noindent Example 3. $\tau(x) = 1 + x_1 + x_2 + 2x_3 + 0.5x_4,
                        \xi(x) = \sign(0.146 + \sin(4  \pi  x_1) + x_2^2)$.

                        \noindent Example 4. 
                        $\tau(x) =  1 + x_1^2 + x_2^2 + 2 x_3^2 + 0.5 x_4^2,
                        \xi(x) = 3.8(0.8 - x_1^2 - x_2^2)$.
                        
                        \noindent Example 5. 
                        $\tau(x) =  1 + x_1^2 + x_2^2 + 2 x_3^2 + 0.5 x_4^2,
                        \xi(x) = \sign(0.8 - x_1^2 - x_2^2)$.

We evaluate the binomial loss and its truncated robust counterpart using exponential, Matérn 3/2, and Gaussian kernels. For brevity, we refer to the binomial loss configurations by their kernel names (e.g., Exponential) and denote the robust variants by appending the suffix ‘-Robust’ (e.g., Exponential-Robust). These methods are collectively referred to as BRWL to distinguish them from three benchmarks: penalized Q-learning (QL) of \citet{qian2011performance}, outcome weighted learning (WSVM) of \citet{zhao2012estimating}, and residual weighted learning (RWL) of \citet{zhou2017residual}.
For the latter two methods, Gaussian kernels are employed.

For each replicate, we generate training and tuning sets of equal size ($n \in \{100, 400\}$) and a test set of size $N_{\text{test}} = 10,000$. The training set is used to estimate the model parameters, the tuning set is used to select optimal hyperparameters, and the test set is used to evaluate the performance of the final model.
Optimal hyperparameters are selected via a grid search over the regularization parameter $\lambda \in \{10^k \mid k = -3, -2 \dots, 3\}$ and kernel bandwidth $\varrho \in \{10^k \mid k = -1, -0.75, \dots, 1\}$. Following \citet{wang2024unified}, the truncation parameter $\sigma$ is tuned in descending order to enhance robustness. We set $\sigma=2$ for Example 2, $\sigma=1$ for Example 3, and select $\sigma
\in \{2, 4, 8\}$ for Examples 4 and 5. 
Performance is assessed using the misclassification error against the true optimal rule $d^*$ and the self-normalized value function estimator: \(\frac{\mathbb{P}_{n}^{*}\left[\mathbb{I}\{A = d(\boldsymbol{X})\} R / \operatorname{Pr}(A)\right]}{\mathbb{P}_{n}^{*}\left[\mathbb{I}\{A = d(\boldsymbol{X})\} / \operatorname{Pr}(A)\right]}\), where $\mathbb{P}_{n}^{*}$ denotes the empirical average over the test set \citep{murphy2001marginal}.

                To assess outlier robustness, we contaminate the training and tuning data at a specified proportion. For each randomly selected outlier $(\bmx_i, a_i, r_i)$, we invert the optimal decision rule by drawing the contaminated response from $R_i \sim N(\tau(\bmx_i) - \xi(\bmx_i) a_i, 1)$.

The simulations are repeated 100 times. We summarize the performance using the mean and standard deviation of the estimated value function and the misclassification rate. Tables~\ref{tab:ex8p50smooth-data} and \ref{tab:ex8p50nonsmooth-data} display the results for Examples 2 and 3 in the moderate-dimensional setting, while the additional results are provided in the supplementary material.
                In Example 2 (smooth), QL performs best in low-dimensional clean settings for both $n=100$ and $n=400$, and dominates all scenarios once $n$ reaches $400$, although robust BRWL with the exponential-kernel remains superior under contamination or higher dimensionality. For Example 3 (nonsmooth), the exponential kernel is generally preferred, particularly in moderate dimensions, though QL leads in low-dimensional cases at $n=400$. In Example 4 (smooth), performance varies by
                dimension: RWL and Gaussian-Robust BRWL outperform other methods in
                low-dimensional settings, whereas the exponential kernel generally dominates moderate dimensions—except for a slight advantage for the Matérn 3/2 kernel at $n=100$ with 5\% contamination. 
Finally, in the nonsmooth case (Example 5), RWL dominates in low-dimensional settings; however, in moderate dimensions, robust BRWL with Matérn or Gaussian kernels achieves optimality at $n = 100$, while the exponential kernel exhibits the best performance at $n = 400$.

                        \section{Data Application}\label{sec:data}
                        We analyze data from the AIDS Clinical Trials Group Study 175 (ACTG175) \citep{hammer1996trial}, a randomized clinical trial assessing antiretroviral regimens for HIV-1 management. Participants were assigned to one of four groups: zidovudine monotherapy $(n=532)$, didanosine monotherapy $(n=561)$, 
                        zidovudine plus didanosine $(n=522)$, or zidovudine plus zalcitabine $(n=524)$. 
                        Our focus is on pairwise comparisons of CD4 cell count changes between the 20-week follow-up and baseline, contrasting zidovudine monotherapy with the other three treatments.

                        The baseline demographic and clinical covariates analyzed include age, gender, race, weight, homosexual activity, intravenous drug use history, symptomatic indicator, antiretroviral history, hemophilia, Karnofsky score, and baseline CD4/CD8 cell counts. Subjects are randomly divided into training $(n=400)$, tuning $(n=400)$, and test sets (remaining). Continuous variables, including the outcome, are standardized based on training set statistics to ensure consistency
                        across the tuning and test data.
We assess robustness by randomly flipping treatment assignments in the training and tuning sets at rates of $0\%, 5\%$, and $10\%$, while leaving the test data unchanged.
                        \begin{table}[htbp!]
    \centering
\begin{threeparttable}
 \caption{Means and standard deviations (in parentheses) of the estimated value functions and the              misclassification rates for Example 2 with $m=50$ covariates.}
    \label{tab:ex8p50smooth-data}
% --- Apply the size command here ---
    \scriptsize
% Define the column types once: 1 Method column (l) and 6 data columns (r)
    \begin{tabular}{l *{6}{r}} 
        \toprule
        % --- TOP HEADER ROW ---
        \multirow{2}{*}{Method} & \multicolumn{2}{c}{0\% Outliers} & \multicolumn{2}{c}{5\% Outliers} & \multicolumn{2}{c}{10\% Outliers} \\%& \multicolumn{2}{c}{[Other Metric]} \\
        \cmidrule(lr){2-3} \cmidrule(lr){4-5} \cmidrule(lr){6-7} %\cmidrule(lr){8-9}
        % --- SECOND HEADER ROW (Metrics) ---
        & Value & Error & Value & Error & Value & Error \\
        \midrule
        
        % -------------------------------------------------------------
        % --- BLOCK 1: SAMPLE SIZE n=100 ---
        \multicolumn{7}{l}{\textbf{Sample Size: $n=100$}} \\
        \cmidrule(lr){1-7} % Horizontal rule spanning all columns
Exponential & 1.463 (0.044) & 0.329 (0.015) & 1.446 (0.086) & 0.334 (0.031) & 1.434 (0.103) & 0.339 (0.039)\\
\Matern 3/2 & 1.430 (0.069) & 0.341 (0.024) & 1.413 (0.096) & 0.347 (0.034) & 1.404 (0.117) & 0.351 (0.043)\\
Gaussian & 1.412 (0.082) & 0.348 (0.030) & 1.392 (0.105) & 0.355 (0.038) & 1.383 (0.128) & 0.359 (0.047)\\
WSVM & 1.296 (0.378) & 0.390 (0.141) & 1.271 (0.394) & 0.400 (0.146) & 1.232 (0.421) & 0.415 (0.157)\\
QL & 1.367 (0.147) & 0.364 (0.054) & 1.322 (0.151) & 0.380 (0.056) & 1.275 (0.182) & 0.397 (0.068)\\
RWL & 1.420 (0.096) & 0.344 (0.034) & 1.407 (0.120) & 0.349 (0.044) & 1.409 (0.119) & 0.349 (0.043)\\
Exponential-Robust & \textbf{1.464} (0.044) & \textbf{0.328} (0.015) & \textbf{1.446} (0.086) & \textbf{0.334} (0.031) & \textbf{1.434} (0.103) & \textbf{0.339} (0.038)\\
\Matern 3/2-Robust & 1.450 (0.064) & 0.333 (0.023) & 1.436 (0.098) & 0.338 (0.035) & 1.425 (0.114) & 0.343 (0.042)\\
Gaussian-Robust & 1.442 (0.075) & 0.337 (0.025) & 1.427 (0.096) & 0.342 (0.034) & 1.417 (0.122) & 0.346 (0.044)\\

        \midrule
        % -------------------------------------------------------------
        % --- BLOCK 2: SAMPLE SIZE n=400 ---
        \multicolumn{7}{l}{\textbf{Sample Size: $n=400$}} \\
        \cmidrule(lr){1-7} 
Exponential & 1.476 (0.028) & \textbf{0.324} (0.006) & 1.473 (0.029) & 0.324 (0.007) & \textbf{1.470} (0.030) & \textbf{0.325} (0.008)\\
\Matern 3/2 & 1.466 (0.034) & 0.327 (0.009) & 1.459 (0.038) & 0.329 (0.012) & 1.456 (0.040) & 0.331 (0.014)\\
Gaussian & 1.456 (0.040) & 0.330 (0.011) & 1.448 (0.047) & 0.333 (0.016) & 1.441 (0.054) & 0.336 (0.019)\\
WSVM & 1.403 (0.264) & 0.350 (0.097) & 1.360 (0.316) & 0.366 (0.116) & 1.343 (0.337) & 0.373 (0.124)\\
QL & 1.463 (0.041) & 0.328 (0.013) & 1.459 (0.043) & 0.330 (0.014) & 1.447 (0.055) & 0.334 (0.019)\\
RWL & 1.466 (0.042) & 0.327 (0.013) & 1.461 (0.047) & 0.329 (0.015) & 1.450 (0.059) & 0.333 (0.021)\\
Exponential-Robust & \textbf{1.476} (0.027) & 0.324 (0.006) & \textbf{1.473} (0.028) & \textbf{0.324} (0.007) & 1.470 (0.030) & 0.326 (0.008)\\
\Matern 3/2-Robust & 1.471 (0.031) & 0.325 (0.009) & 1.467 (0.033) & 0.327 (0.010) & 1.466 (0.035) & 0.327 (0.011)\\
Gaussian-Robust & 1.469 (0.032) & 0.326 (0.010) & 1.466 (0.034) & 0.327 (0.012) & 1.462 (0.040) & 0.329 (0.014)\\

\bottomrule
\end{tabular}
\begin{tablenotes}
      \small
       \item NOTE: For each scenario, the maximum value function   and minimum                misclassification rate are highlighted in bold. 
\end{tablenotes}
\end{threeparttable}
\end{table}

                        \newpage
\begin{table}[htbp!]
    \centering
    \begin{threeparttable}
 \caption{Means and standard deviations (in parentheses) of the estimated value functions and               the misclassification rates for Example 3 with $m=50$ covariates.}
    \label{tab:ex8p50nonsmooth-data}
% --- Apply the size command here ---
    \scriptsize
% Define the column types once: 1 Method column (l) and 6 data columns (r)
    \begin{tabular}{l *{6}{r}} 
        \toprule
        % --- TOP HEADER ROW ---
        \multirow{2}{*}{Method} & \multicolumn{2}{c}{0\% Outliers} & \multicolumn{2}{c}{5\% Outliers} & \multicolumn{2}{c}{10\% Outliers} \\%& \multicolumn{2}{c}{[Other Metric]} \\
        \cmidrule(lr){2-3} \cmidrule(lr){4-5} \cmidrule(lr){6-7} %\cmidrule(lr){8-9}
        % --- SECOND HEADER ROW (Metrics) ---
        & Value & Error & Value & Error & Value & Error \\
        \midrule
        
        % -------------------------------------------------------------
        % --- BLOCK 1: SAMPLE SIZE n=100 ---
        \multicolumn{7}{l}{\textbf{Sample Size: $n=100$}} \\
        \cmidrule(lr){1-7} % Horizontal rule spanning all columns
Exponential & \textbf{1.334} (0.058) & \textbf{0.335} (0.026) & 1.322 (0.075) & 0.340 (0.034) & 1.306 (0.099) & 0.349 (0.048)\\
\Matern 3/2 & 1.307 (0.073) & 0.349 (0.034) & 1.296 (0.088) & 0.354 (0.041) & 1.284 (0.103) & 0.360 (0.050)\\
Gaussian & 1.284 (0.088) & 0.360 (0.042) & 1.281 (0.097) & 0.361 (0.045) & 1.268 (0.110) & 0.368 (0.054)\\
WSVM & 1.180 (0.309) & 0.410 (0.153) & 1.149 (0.326) & 0.426 (0.162) & 1.124 (0.334) & 0.438 (0.167)\\
QL & 1.188 (0.130) & 0.407 (0.065) & 1.153 (0.133) & 0.426 (0.066) & 1.123 (0.139) & 0.441 (0.071)\\
RWL & 1.292 (0.096) & 0.356 (0.047) & 1.280 (0.102) & 0.361 (0.048) & 1.276 (0.112) & 0.364 (0.056)\\
Exponential-Robust & 1.333 (0.059) & 0.336 (0.026) & \textbf{1.323} (0.075) & \textbf{0.340} (0.034) & \textbf{1.309} (0.098) & \textbf{0.347} (0.048)\\
\Matern 3/2-Robust & 1.323 (0.069) & 0.341 (0.031) & 1.302 (0.095) & 0.351 (0.045) & 1.301 (0.099) & 0.352 (0.049)\\
Gaussian-Robust & 1.311 (0.081) & 0.346 (0.038) & 1.301 (0.096) & 0.351 (0.046) & 1.293 (0.105) & 0.356 (0.051)\\

\midrule
        % -------------------------------------------------------------
        % --- BLOCK 2: SAMPLE SIZE n=400 ---
        \multicolumn{7}{l}{\textbf{Sample Size: $n=400$}} \\
        \cmidrule(lr){1-7} 

\midrule

Exponential & \textbf{1.347} (0.036) & \textbf{0.328} (0.012) & 1.342 (0.038) & 0.330 (0.014) & 1.338 (0.043) & 0.333 (0.018)\\
\Matern 3/2 & 1.331 (0.045) & 0.336 (0.019) & 1.329 (0.046) & 0.337 (0.021) & 1.325 (0.054) & 0.339 (0.024)\\
Gaussian & 1.322 (0.048) & 0.341 (0.022) & 1.321 (0.056) & 0.341 (0.025) & 1.308 (0.067) & 0.348 (0.031)\\
WSVM & 1.248 (0.261) & 0.376 (0.128) & 1.242 (0.267) & 0.380 (0.131) & 1.228 (0.278) & 0.387 (0.137)\\
QL & 1.331 (0.056) & 0.337 (0.024) & 1.319 (0.065) & 0.343 (0.031) & 1.304 (0.073) & 0.349 (0.036)\\
RWL & 1.326 (0.056) & 0.339 (0.024) & 1.327 (0.057) & 0.338 (0.026) & 1.317 (0.066) & 0.343 (0.031)\\
Exponential-Robust & 1.346 (0.034) & 0.329 (0.011) & \textbf{1.345} (0.036) & \textbf{0.329} (0.012) & \textbf{1.343} (0.038) & \textbf{0.331} (0.015)\\
\Matern 3/2-Robust & 1.338 (0.046) & 0.333 (0.018) & 1.337 (0.040) & 0.332 (0.017) & 1.329 (0.053) & 0.337 (0.023)\\
Gaussian-Robust & 1.340 (0.041) & 0.332 (0.018) & 1.333 (0.050) & 0.335 (0.022) & 1.335 (0.047) & 0.334 (0.021)\\

\bottomrule
\end{tabular}
\begin{tablenotes}
      \small
       \item NOTE: For each scenario, the maximum value function   and minimum                misclassification rate are highlighted in bold. 
\end{tablenotes}
\end{threeparttable}
\end{table}

We consider the residual-based framework in Section~\ref{sec:negrewards} using the binomial loss and its robust truncated version (with truncation parameter $\sigma = 1$), applied across exponential, Matérn 3/2, and Gaussian kernels, which are jointly referred to as BRWL. We employ tuning procedures for the kernel width and regularization parameters consistent with those described in the simulation study (Section~\ref{sec:sim}).
A similar analysis is conducted for the competing methods to ensure a fair comparison. Results, including means and standard deviations of estimated value functions from $100$ replicates, are presented in Table~\ref{tab:thirdtype} and supplementary material. The findings indicate that the BRWL methods, particularly the robust version, outperform or are comparable to competing approaches and demonstrate greater resilience to outliers.
\begin{table}[H]
\centering
\caption{Means and standard deviations (in parenthesis) of the estimated value functions for the ACTG175 trial: zidovudine monotherapy vs. didanosine monotherapy.}
\label{tab:thirdtype}
%\resizebox{\linewidth}{!}{
\begin{tabular}{lrrrr}
\toprule
Method & 0\% Outliers & 5\% Outliers & 10\% Outliers \\
\midrule

 Exponential & 0.179 (0.093) & 0.177 (0.091) & 0.169 (0.093)\\
 Matern 3/2 & 0.176 (0.090) & 0.172 (0.094) & 0.166 (0.093)\\
 Gaussian & 0.169 (0.094) & 0.170 (0.098) & 0.163 (0.098)\\
 WSVM & 0.136 (0.115) & 0.125 (0.124) & 0.110 (0.126)\\
 QL & 0.167 (0.089) & 0.158 (0.095) & 0.147 (0.107)\\
 RWL & 0.174 (0.091) & 0.166 (0.093) & 0.157 (0.094)\\
 Exponential-Robust & \textbf{0.181} (0.090) & \textbf{0.178} (0.091) & \textbf{0.175} (0.089)\\
 Matern 3/2-Robust & 0.181 (0.088) & 0.172 (0.091) & 0.171 (0.091)\\
 Gaussian-Robust & 0.176 (0.090) & 0.170 (0.089) & 0.166 (0.092)\\

\bottomrule
\end{tabular}
%}
\end{table}

                \section{Conclusions}\label{sec:dis}
In this work, we studied statistical properties of OWL using a nonnegative surrogate loss $T$ that targets the 0--1 misclassification loss. Assuming the minimum is attainable by the intended algorithms, we examined three sources of error: (i) estimation error due to finite samples, (ii) approximation error arising from the restricted function class $\mathcal{H}$, and (iii) approximation error incurred by replacing the 0--1 loss with a surrogate loss. 

For the third source of error, we presented a universal inequality that upper bounds the population 0--1 risk by the population $T$-risk. This relationship is characterized by the $\Psi$-transform, a convexified variational mapping of the original loss. We established conditions under which $T$-risk consistency implies Bayes-risk consistency, where $T$-risk consistency refers to the sequential convergence of population $T$-risks to the global minimum over all measurable classifiers. A key
requirement is the positive definiteness of the $\Psi$-transform. Although $\Psi$-positive definiteness guarantees policy-calibration, the converse may not hold, unlike classification-calibration, which is equivalent to $\psi$-positive definiteness. Nonetheless, many commonly used loss functions satisfy $\Psi$-positive definiteness, enabling inversion of the corresponding risk bound. For convex losses $T$, the analysis simplifies substantially: policy-calibration reduces to verifying a
negative derivative at the origin, and convexity eliminates the need for convexification in constructing $\Psi$, permitting direct closed-form expressions. For certain nonconvex losses, we showed that the $\Psi$-transform is linear and applied this result to robust binomial losses in the CC family.

For the first source of error, estimation error, we employed kernel methods to estimate optimal ITRs and established upper bounds for both convex and bounded nonconvex losses. For the second source of error, approximation error, we assumed that the target function lies in a Sobolev space. Under \Matern\ kernels, we obtained convergence rates for both convex and bounded nonconvex loss functions, with the added advantage that these kernel methods adapt automatically to the unknown smoothness of the target function. Under geometric noise assumptions, we further derived convergence rates for both loss types, with specific applications to Gaussian kernels. In general, convex loss functions achieve faster convergence rates.

Extending the OWL framework to RWL for general loss functions remains a topic for future work. Another promising direction is to obtain sharper convergence rates for strictly convex losses \citep{bartlett2006convexity}, which can improve upon rates derived from standard uniform convergence arguments. Moreover, even faster rates may be possible under complete data separation \citep{zhao2012estimating}.

The transition from two-arm to multi-arm trials for OWL has been studied \citep{zhang2020multicategory}. Future research could focus on extending policy-calibration and the $\Psi$-transform through multicategory classification using specialized loss functions like the softmax \citep{hastie2009elements} or decomposition methods such as one-versus-one or one-versus-all.

Multi‑stage dynamic treatment regimes are often more clinically relevant for chronic disease management than single‑stage rules, as they adapt decisions to a patient’s evolving health status. Although \citet{zhao2015new} extended OWL to this setting, extending our proposed methods to multi‑stage regimes remains an avenue for future research.

\phantomsection\label{supplementary-material}
\bigskip

\begin{center}

{\large\bf SUPPLEMENTARY MATERIAL}

\end{center}

%\begin{description}
This supplement contains three sections. Section 1 provides the proofs for the theoretical results; Section 2 details the derivation of Table~\ref{tab:pc1}; and Section 3 contains extended simulation results and additional results from the ACTG175 trial.
%\item[Title:]
%Brief description. (file type)
%\item[R-package for MYNEW routine:]
%R-package MYNEW containing code to perform the diagnostic methods
%described in the article. The package also contains all datasets used as
%examples in the article. (GNU zipped tar file)
%\item[HIV data set:]
%Data set used in the illustration of MYNEW method in
%Section~\ref{sec-verify} (.txt file).
%\end{description}
\section*{Data Availability Statement}
R code and an analysis of the ACTG175 trial data, including an R package and the associated data, will be made available on CRAN upon publication of this manuscript.
\section*{Disclosure Statement}
The author reports there are no competing interests to declare.
\section*{Acknowledgment}
Google Gemini 2.5 and 3.0 were used during manuscript preparation for idea exploration, language refinement, and coding support. All AI‑assisted content was reviewed, verified, and edited by the author, who takes full responsibility for the final manuscript.

\begin{spacing}{1}
\bibliography{../wangres}
\end{spacing}
\includepdf[pages=-]{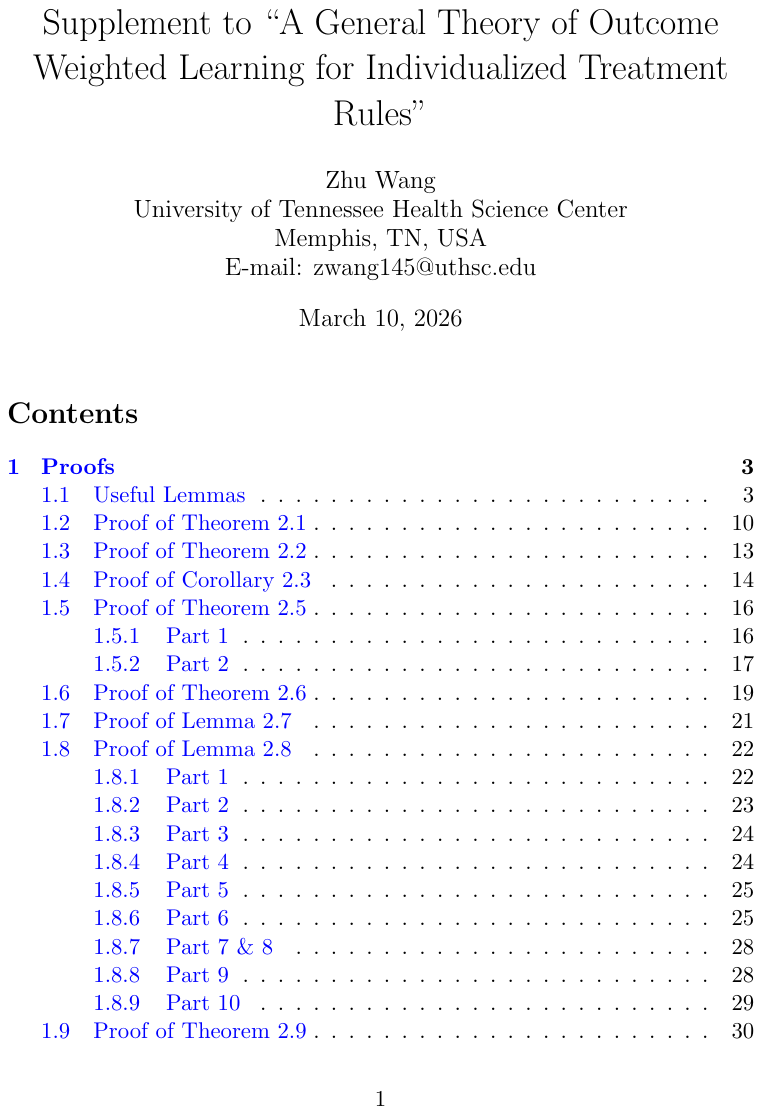}
\end{document}